\documentclass[%
aip,
amsmath,amssymb,
reprint,%
]{revtex4-2}

\usepackage{graphicx}
\usepackage{subcaption}
\usepackage[
binary-units,
]{siunitx}
\usepackage{dcolumn}
\usepackage{bm,bbm}
\usepackage[mathscr]{eucal}

\usepackage[utf8]{inputenc}
\usepackage[T1]{fontenc}
\usepackage{mathptmx}
\usepackage{etoolbox}
\usepackage{xcolor}
\usepackage{booktabs}
\usepackage{diagbox}

\newtheorem{thm}{Theorem}

\makeatletter
\def\@email#1#2{%
\endgroup
\patchcmd{\titleblock@produce}
{\frontmatter@RRAPformat}
{\frontmatter@RRAPformat{\produce@RRAP{*#1\href{mailto:#2}{#2}}}\frontmatter@RRAPformat}
{}{}
}%
\makeatother
\begin{document}

\preprint{AIP/123-QED}

\title[Statistics Entropy Ordinal Patterns]{Statistical Properties of the Entropy from Ordinal Patterns}
\author{E.\ T.\ C.\ Chagas}
\affiliation{Departamento de Ci\^encia da Computa\c c\~ao, Universidade Federal de Minas Gerais, Belo Horizonte, Brazil}
\email{eduarda.chagas@dcc.ufmg.br}

\author{A.\ C.\ Frery}%
\affiliation{School of Mathematics and Statistics, Victoria University of Wellington, Wellington, New Zealand}

\author{J.\ Gambini}
\affiliation{Instituto Tecnol\'ogico de Buenos Aires - Universidad Nacional de Tres de Febrero, Buenos Aires, Rep\'ublica Argentina}

\author{M.\ M.\ Lucini}
\affiliation{ 
Universidad Nacional del Nordeste,
Corrientes, and CONICET, Buenos Aires, Rep\'ublica Argentina}

\author{H.\ S.\ Ramos}
\affiliation{Departamento de Ci\^encia da Computa\c c\~ao, Universidade Federal de Minas Gerais, Belo Horizonte, Brazil}

\author{A.\ A.\ Rey}
\affiliation{Universidad Tecnol\'ogica Nacional, Facultad Regional Buenos Aires, Buenos Aires, Rep\'ublica Argentina}%

\date{\today}

\begin{abstract}
The ultimate purpose of the statistical analysis of ordinal patterns is to characterize the distribution of the features they induce.
In particular, knowing the joint distribution of the pair Entropy-Statistical Complexity for a large class of time series models would allow statistical tests that are unavailable to date.
Working in this direction, we characterize the asymptotic distribution of the empirical Shannon's Entropy for any model under which the true normalized Entropy is neither zero nor one.
We obtain the asymptotic distribution from the Central Limit Theorem (assuming large time series), the Multivariate Delta Method, and a third-order correction of its mean value.
We discuss the applicability of other results (exact, first-, and second-order corrections) regarding their accuracy and numerical stability.
Within a general framework for building test statistics about Shannon's Entropy, we present a bilateral test that verifies if there is enough evidence to reject the hypothesis that two signals produce ordinal patterns with the same Shannon's Entropy.
We applied this bilateral test to the daily maximum temperature time series from three cities (Dublin, Edinburgh, and Miami) and obtained sensible results.
\end{abstract}

\maketitle

\begin{quotation}
We present results about the statistical properties of Ordinal Patterns' Shannon Entropy: exact and approximate first-, second-, and third-order moments and its asymptotic distribution.
We discuss their limitations and propose a new model.
We present applications to test the hypothesis that two time series produce ordinal patterns with the same Shannon entropy.
\end{quotation}

\section{Introduction}

The analysis of signals throughout their ordinal patterns has received much attention since it was proposed by \citet{PermutationEntropyaNaturalComplexityMeasureforTimeSeries}.
This approach is appealing, among other virtues, for its ability to translate analysis into a graphical depiction: the signal is represented as a point in the Entropy-Complexity plane ($H\times C$), a closed two-dimensional manifold.

There is a vast literature of successful applications of this kind of analysis, among them:
The ability to distinguish noise from chaos\cite{DistinguishingNoisefromChaos},
the analysis of economic markets and biomedical systems\cite{PermutationEntropyandItsMainBiomedicalandEconophysicsApplicationsaReview},
fault diagnosis of rotating machinery\cite{FaultDiagnosisofRotatingMachineryaReviewandBibliometricAnalysis},
and image texture analysis\cite{AnalysisandClassificationofSARTexturesUsingInformationTheory,DiscriminatingImageTextureswiththeMultiscaleTwoDimensionalComplexityEntropyCausalityPlane,ComplexityEntropyCausalityPlaneAsaComplexityMeasureforTwoDimensionalPatterns}.

\citet{WhiteNoiseTestfromOrdinalPatternsintheEntropyComplexityPlane} pointed out a common trait of such applications: the lack of a theoretical background when performing comparisons among signals and when contrasting a signal with a hypothesized model.
The authors proposed a test for the white noise model using empirical evidence.
Although this approach proved useful, extending it to other situations requires extensive simulation experiments and data analysis.

The ultimate goal in a statistical approach to the Bandt and Pompe methodology is the exact specification of the distribution of the coordinates of points in $H\times C$ for a large class of time series models.
Our purpose is delimited:
In this work, we present exact, approximate, and asymptotic results of the distribution of Shannon's Entropy when the patterns that result from the Bandt and Pompe symbolization follow a Multinomial distribution.
The properties of such a distribution that we employ later are recalled in Section~\ref{Sec:OP_and_Multinomial}.

The exact and approximate results we present in Section~\ref{Sec:EntropyProperties} appeared in the works by
\citet{Basharin_59}, 
by \citet{ATestforComparingDiversitiesBasedontheShannonFormula}, 
and by \citet{SomeMomentsofanEstimateofShannonsMeasureofInformation}.
These works have received little attention from the community that uses the permutation entropy approach.

We discuss the applicability of these expressions in Section~\ref{Sec:Numerical}, stressing that the exact ones are of limited use in practice with high-standard numerical platforms currently available.
We obtain the asymptotic distribution of Shannon's Entropy by using the properties of the Multinomial distribution, the Central Limit Theorem, and the Multivariate Delta method  (Section~\ref{Sec:Asymptotic}).
We propose a better approximation by plugging a third-order correction of the mean into its asymptotic expression (Eq.~\eqref{Eq:ProposedMeanH}).

We also provide a general framework for hypotheses testing, including the possibility of contrasting series of different lengths and embedding dimensions.
We explicitly derive tests statistics for the null hypothesis that two time series produce the same number of symbols, and we apply these tests to climatology data.

\section{Ordinal Patterns and the Multinomial Distribution}\label{Sec:OP_and_Multinomial}

Let \(\bm x=(x_1,x_2,\dots,x_{n+D-1})\) be a real-valued time series of length \(n+D-1\) without ties. 
We compute \(\pi_1,\pi_2,\dots, \pi_{n}\) symbols from
subsequences of embedding dimension \(D\), for instance
\(\pi_j=\text{BP}(x_j,x_{j+1},\dots,x_{j+D-1 })\) where \(\text{BP}\) is
the Bandt-Pompe symbolization. 
There are \(D!\) possible symbols:
\(\pi_j\in \bm \pi =\{\pi^1, \pi^2, \dots, \pi^{D!}\}\). 
We form the
histogram of proportions \(\bm h = (h_1, h_2, \dots, h_{D!})\) in which
the bin \(h_\ell\) is the proportion of symbols of type \(\pi^{\ell}\)
of the total \(n\) symbols.
We will model those symbols as a $k$ dimensional random vector with a multinomial distribution in which $k=D!$.

Consider a series of \(n\) independent trials in which only one of \(k\) mutually exclusive events \(\pi^1, \pi^2, \dots, \pi^k\) is
observed with probability \(p_1,p_2,\dots,p_k\), respectively, such that $p_\ell\geq 0$ and
\(\sum_{\ell=1}^k p_\ell=1\).
Let \(\bm N = (N_1,N_2,\dots,N_k)\) be the
vector of random variables that count the number of occurrences of the events
\(\pi^1, \pi^2, \dots, \pi^k\) in the \(n\) trials, with
\(\sum_{\ell=1}^k N_\ell = n\). Then, the joint distribution of \(\bm N\)
is \begin{equation}
\Pr\big(\bm N = (n_1,n_2,\dots,n_k)\big) = n! \prod_{\ell=1}^k \frac{p_\ell^{n_\ell}}{n_\ell!},
\label{eq:DistMultinomial}
\end{equation} where \(n_\ell\geq 0\) and \(\sum_{\ell=1}^k n_\ell = n\).
We denote this situation as \(\bm N\sim \text{Mult}(n,\bm p)\), with \(\bm p = (p_1,p_2,\dots,p_k)\).

As per the conditions under which the Bandt and Pompe technique is used, we
require that \(n\gg k\).
The common practice is $k=3!$, $4!$, $5!$ or $6!$, and \(n\ge 100 k\).

Consider the random vector \(\bm N\sim \text{Mult}(n,\bm p)\). 
Its main moments are: \begin{align}
\operatorname{E}(N_{\ell}) &= n p_{\ell},\\
\operatorname{Var}(N_{\ell}) &= n p_{\ell}(1-p_{\ell}), \\
\operatorname{Cov}(N_{\ell},N_j) &= -n p_{\ell} p_j, \text{ and}\\
\varrho(N_{\ell},N_j) &= \sqrt{\frac{p_{\ell} p_j}{(1-p_{\ell})(1-p_j)}},
\end{align}
for every $1\leq \ell,j\leq k$.
The book by \citet[Section~11.18]{johnson_discrete}, is a comprehensive treatise on this topic.

The maximum likelihood (ML) estimator of \(p_\ell\) is the relative frequency
\(\widehat{p}_\ell = N_\ell/n\), $1\leq \ell\leq k$, and the distribution of
\(n\widehat{\bm p}\) is \(\text{Mult}(n,\bm p)\).
The properties of ML estimators grant that if $\widehat{\bm p}$ is the ML estimator of $\bm p$, then for any function $g(\bm p)$, the ML estimator of $g(\bm p)$, namely $\widehat{g}(\bm p)$, is $g(\widehat{\bm p})$; cf.\ \citet[Theorem~7.2.10]{CasellaBergerStatisticalInference}.
We will use this result to obtain the asymptotic distribution of Shannon's Entropy.

Let \(\bm X_n = (X_{1n}, X_{2n}, \dots ,X_{kn})\) be a sequence of independent  and identically distributed random vectors,  with  \(\text{Mult}(n,\bm p)\)  distribution. 
If \(\widehat{\bm{p}}\) is the vector of sample proportions and $\bm{Y}_n = \sqrt{n}(\widehat{\bm{p}}-\bm{p})$, then 
\begin{align}
\operatorname{E}(\bm{Y}_n) &= 0,\\
\operatorname{Cov}(\bm{Y}_n) &= \textbf{D}_{\bm p}-\bm{p}\bm{p}^{\text{T}},
\end{align} 
where $\textbf{D}_{\bm p} = \text{Diag}(p_1, p_2, \dots, p_k)$, and the superscript $\text{T}$ denotes transposition.
\citet[Theorem~A.2.1]{mukhopadhyay2016complex} states that
\begin{equation}
\bm{Y}_n \xrightarrow{\mathcal{D}} \mathcal{N}\big(\bm{0}, \textbf{D}_{\bm p}-\bm{p}\bm{p}^{\text{T}}\big).
\label{eq:YnDist}
\end{equation}
This asymptotic distribution is the basis of our forthcoming derivations.

\section{The Entropy and its properties}\label{Sec:EntropyProperties}

This section presents the main results about $H(\widehat{\bm p})$ under the Multinomial model.
Regarding its mean and variance, we recall exact expressions, first, second and third-order approximations. Finally, its asymptotic distribution is presented.

The Shannon's Entropy of a Multinomial-distributed random variable is
\begin{equation}
H(\bm p) = -\sum_{\ell=1}^k p_\ell \ln p_\ell,
\label{eq:EntropyMultinomial}
\end{equation} 
which is bounded between $0$ and $\ln k$.
The minimum is attained when $p_\ell=1$ for some $1 \leq l \leq k$ and $p_j=0$ for every $j\neq \ell$,
while the expression is maximized by \(p_\ell=1/k\) for every $1 \le \ell \le k$.
In the following, we will consider only probability vectors $\bm p$ that differ from these two extreme points.

We are interested in
the statistical properties of \(H(\bm p)\) when it is indexed by \(\widehat{\bm p}=(\widehat{p}_1, \widehat{p}_2, \dots, \widehat{p}_k)\),
the ML estimator of \(\bm p\). 
Our problem then becomes in
finding the distribution of \begin{align}
H(\widehat{\bm p}) & = -\sum_{\ell=1}^k \widehat{p}_\ell \ln \widehat{p}_\ell \label{eq:Hhatpell}\\
& = - \sum_{\ell=1}^k \frac{N_\ell}n \ln \frac{N_\ell}n \nonumber \\
& = \ln n - \frac{1}n \sum_{\ell=1}^k N_\ell \ln N_\ell \label{eq:HhatNell},
\end{align} 
under
\(\bm N = (N_1,N_2,\dots,N_k)\sim\text{Mult}(n,\bm p)\).

\subsection{Exact Mean and Variance}\label{Sec:Exact}

\citet{SomeMomentsofanEstimateofShannonsMeasureofInformation}
found the expressions for the exact first and second-order moments of~\eqref{eq:Hhatpell}.
They are given by:
\begin{widetext}
	\begin{align}
		\operatorname{E}\big( H(\widehat{\bm p}) \big) & =
		\ln n - \sum_{j=1}^{n-1} \binom{n-1}{n-j} \ln(n-j+1) 
		\sum_{\ell=1}^{k} p_{\ell}^{n-j+1}(1-p_\ell)^{j-1},
		\label{Eq:ExactMeanH}\\
		\intertext{and, denoting the integer part of $x\in\mathbbm R$ as $[x]$,}
		\operatorname{Var} \big( H(\widehat{\bm p}) \big) & =
		\sum_{a=0}^{n-2} \binom{n-1}{a} \sum_{\ell=1}^{k} p_\ell^{n-a} (1-p_\ell)^a
		\left[ \sum_{b=a+1}^{n-1} \binom{n-1}{b} \sum_{\ell=1}^{k} p_\ell^{n-b}(1-p_\ell)^b \Big( \ln\frac{n-a}{n-b} \Big)^2
		\right] \nonumber \\
		& \mbox{}-
		\frac{n-1}{n} \sum_{b=0}^{n-3} \binom{n-2}{b}
		\left[
		\sum_{a=0}^{[(n-b-2)/2]} \binom{n-b-2}{a} \mathop{\sum\sum}\limits_{\ell \neq j} 
		p_\ell^{n-a-b-1} (1-p_\ell)^{a+1} (1-p_\ell p_j)^b \Big( \ln\frac{n-a-b-1}{a+1} \Big)^2
		\right].\label{Eq:ExactVarH}
	\end{align} 
\end{widetext}

\subsection{First-order approximation}\label{Sec:FirstOrder}

\citet{Basharin_59} provided some of the first results about the properties of $H(\widehat{\bm p})$.
This author found first-order approximations for its expected value and variance, whose expressions are given by:
\begin{align}
	\check{\operatorname{E}}\big(H(\widehat{\bm p})\big) &= H(\bm p) - \frac{k-1}{2n \ln 2} ,\label{Eq:FirstOrderMean} \intertext{ and}
	\check{\operatorname{Var}}\big(H(\widehat{\bm p})\big) &= \frac{1}{n} \Big[\sum_{\ell=1}^{k} p_\ell \frac{\ln^2 p_\ell}{\ln^2 2} - H^2(\bm p)\Big].\label{Eq:FirstOrderVariance}
\end{align}
Notice that Eq.~\eqref{Eq:FirstOrderMean} can be used to build a first-order bias corrected estimator for $H(\bm p)$.
\citet{Basharin_59} also proved that $H(\widehat{\bm p})$ is consistent ($H(\widehat{\bm p})\to H({\bm p})$ when $n\to\infty$) and asymptotically Normal.

\subsection{Third-order approximation}\label{Sec:SecondOrder}
\citet{ATestforComparingDiversitiesBasedontheShannonFormula}  presented 
third-order
approximate expressions by expanding~\eqref{Eq:ExactMeanH} and~\eqref{Eq:ExactVarH} in series and retaining only the initial terms.
With this, we have:
\begin{multline}
	\widetilde{\operatorname{E}}\big( H(\widehat{\bm p}) \big)  =
	H(\bm p) - \frac{k-1}{2n} + \frac{1-\sum_{\ell=1}^k p_\ell^{-1}}{12 n^2} + \\
	\frac{\sum_{\ell=1}^k (p_\ell^{-1}-p_\ell^{-2})}{12 n^3}
	\label{Eq:ApproxMeanH},	
\end{multline}
and
\begin{multline}
	\widetilde{\operatorname{Var}} \big( H(\widehat{\bm p}) \big)  = 
	\frac{\sum_{\ell=1}^k p_\ell \ln^2p_\ell - \big(\sum_{\ell=1}^k p_\ell \ln p_\ell\big)^2}{n} +
	\frac{k-1}{2 n^2} + \\
	\frac{\sum_{\ell=1}^k p_\ell^{-1} - \sum_{\ell=1}^k p_\ell^{-1}\ln p_\ell + \sum_{\ell=1}^k p_\ell^{-1} \sum_{\ell=1}^k p_\ell\ln p_\ell -1}{6 n^3}.
	\label{Eq:ApproxVarH}	
\end{multline}
Notice that dropping the last term in these expressions yields second-order approximations for the mean and variance, respectively. 
Exact expressions should be better than approximations, but we will see in Section~\ref{Sec:Numerical} that the first ones have limited application due to the numerical instabilities they incur. 
On the contrary, all approximations are numerically stable.

\citet{SomeMomentsofanEstimateofShannonsMeasureofInformation}
briefly discussed how to obtain higher-order moments that would be
useful for computing the skewness and kurtosis.

\subsection{Asymptotic distribution}\label{Sec:Asymptotic}

We recall the following theorems known as the Delta Method and its multivariate version. For their proofs, we refer to~\citet{lehmann2006theory}.

\begin{thm} \label{thm:delta_method}
	Let $X_n$ be a sequence of independent and identically distributed random variables such that $\sqrt{n}(X_n-\theta)$ converges in distribution to a $\mathcal{N}(0,\sigma^2)$ law. 
	Consider the transformation $h(X_n)$ such that $h'(\theta)$ exists and does not vanish.
	Then $\sqrt{n}\big[h(X_n)-h(\theta)\big]$ converges in distribution to a $\mathcal{N}\big(0,\sigma^2[h'(\theta)]^2\big)$ law.
\end{thm}

\begin{thm} \label{thm:delta_method_multi}
	Let $\bm X_n = (X_{1n}, X_{2n}, \dots ,X_{kn})$ be a sequence of independent and identically distributed vectors of random variables such that $\sqrt{n}(X_{1n}-\theta_1, X_{2n}-\theta_2, \dots , X_{kn}-\theta_k)$ converges in distribution to the multivariate Normal law $\mathcal{N}_n(\bm 0,\Sigma)$ where $\Sigma$ is the covariance matrix. 
	Suppose that $h_1, h_2, \dots , h_k$ are real-functions continuously differentiable in a neighborhood of the parameter point $\bm{\theta} = (\theta_1, \theta_2, \dots , \theta_k)$ and such that the matrix of partial derivatives $B = (\partial h_i/\partial \theta_j)_{i,j=1}^k$ is non-singular in the mentioned neighborhood. Then, the following convergence in distribution holds
	\begin{multline*}
		\sqrt{n}\big[h_1(\bm{X}_n)-h_1(\bm{\theta}), h_2(\bm{X}_n)-h_2(\bm{\theta}), \dots, h_k(\bm{X}_n)-h_k(\bm{\theta}) \big]\\ \xrightarrow{\mathcal{D}} \mathcal{N}\big(\bm{0}, B \Sigma B^{\text{T}}\big).
	\end{multline*}
\end{thm}

For our case of interest \(\bm N \sim \text{Mult}(n,\bm p)\), the covariance matrix of Eq.~\eqref{eq:YnDist} is
\begin{equation} 
	\big(\textbf{D}_{\bm p} -\bm{p}\bm{p}^{\text{T}}\big)_{\ell j} = \begin{cases}
		p_\ell (1-p_\ell ) & \text{if } \ell  = j, \\
		p_\ell p_j & \text{if } \ell  \ne j.
	\end{cases}
	\label{eq:CovMulti}
\end{equation}
$1 \leq l,j,\leq k$.

In order to apply the Delta Method using Theorem~\ref{thm:delta_method_multi} to Shannon's Entropy defined in Eq.~\eqref{eq:Hhatpell}, we use the functions
\begin{equation}
	h_\ell (p_1, p_2, \dots , p_k) = p_\ell  \ln p_\ell , \label{eq:hS_delta}
\end{equation}
which verify that
\begin{equation}
	\frac{\partial h_\ell }{\partial p_j} = 
	\begin{cases}
		\ln p_\ell  + 1 & \text{if } \ell =j, \\
		0			& \text{otherwise}.
	\end{cases}
\end{equation}
$1 \leq \ell,j,\leq k$.
Hence, the covariance matrix of the multivariate Normal limit distribution 
$\Sigma_{\bm p} = (\partial h_\ell /\partial p_j) (\textbf{D}_{\bm p} -\bm{p}\bm{p}^{\text{T}}) (\partial h_\ell /\partial p_j)^{\text{T}}$ is of the form
\begin{equation} 
	(\Sigma_{\bm p})_{\ell j} = 
	\begin{cases}
		(p_\ell -p_\ell ^2)(\ln p_\ell  + 1)^2 & \text{if }\ell  = j, \\
		-p_\ell  p_j (\ln p_\ell  + 1) (\ln p_j + 1) & \text{otherwise}. \\
	\end{cases} \label{eq:cov_deltaS}
\end{equation}
$1 \leq l,j,\leq k$.
Therefore, we conclude that
\begin{multline} \label{multi_cov_dist}
	\sqrt{n}\big[h_1(\widehat{p}_1) - h_1 (p_1), h_2(\widehat{p}_2) - h_2(p_2), \dots, h_{k}\widehat{p}_{k} - h_{k}(p_{k}) \big]\\ \xrightarrow{\mathcal{D}} \mathcal{N}(\bm{0}, \Sigma_{\bm p}).
\end{multline}
An equivalent expression is:
\begin{equation}
	\sqrt{n}\big[h_1(\widehat{p}_1), h_2(\widehat{p}_2), \dots, h_{k}(\widehat{p}_{k}) \big] \\ \xrightarrow{\mathcal{D}} \mathcal{N}\left(\sqrt{n}\begin{pmatrix}
		h_1(p_1) \\
		h_2(p_2) \\
		\vdots \\
		h_{k}(p_{k})
	\end{pmatrix}, \Sigma_{\bm p}\right).
	\label{Eq:TCLh}
\end{equation}

For a random vector  $\bm Y$ such that $\sqrt{n} \bm Y \xrightarrow{\mathcal{D}}\mathcal{N} (\sqrt{n} \bm \mu, \Sigma)$, it can be proved that $\operatorname{E}(\sqrt{n} \bm Y) \rightarrow \sqrt{n} \bm \mu$ and that $\operatorname{Var}(\sqrt{n} \bm Y) \rightarrow \Sigma$. 
Provided well-known properties, it holds that $\operatorname{E}(\bm Y) \rightarrow \bm \mu$ and $\operatorname{Var}(\bm Y) \rightarrow 1/n \Sigma$. Applying this to Eq.~\eqref{Eq:TCLh}, 
\begin{equation}
	\big[h_1(\widehat{p}_1), h_2(\widehat{p}_2), \dots, h_{k}(\widehat{p}_{k}) \big]\\ \xrightarrow{\mathcal{D}} \mathcal{N}\left(\begin{pmatrix}
		h_1(p_1) \\
		h_2(p_2) \\
		\vdots\\
		h_{k}(p_{k})
	\end{pmatrix}, \frac{1}{n} \Sigma_{\bm p}\right).
	\label{Eq:TCLhmodified}
\end{equation}

We now use Eq.~\eqref{Eq:TCLhmodified} and the fact that the Shannon's Entropy is a linear combination of the functions $\{h_1(p_1), h_2(p_2), \dots, h_{k}(p_{k})\}$.

Let $\bm Z\sim\mathcal N(\bm \mu, \Sigma)$ be a multivariate Gaussian $k$ dimensional vector,  with $\bm \mu \in \mathbbm R^k$ and  $\Sigma=(\sigma_{\ell ,j}), 1\leq \ell,j \leq k$.  Let  $W=\bm a^{\text{T}} \bm Z$ be a linear combination of the $\bm Z$ elements, with $\bm a\in \mathbbm R^k$. Thus, $W$  is $\mathcal N\big(\bm a^{\text{T}}\bm \mu , \sum_{\ell =1}^k a_\ell ^2 \sigma_{\ell ,\ell } + 2 \sum_{\ell =1}^{k-1}\sum_{j=\ell +1}^{k} a_\ell  a_j \sigma_{\ell ,j}\big)$ distributed (see~\citet{lehmann2006theory}).  By using the limit distribution presented in Eq.~\eqref{Eq:TCLhmodified} and $\bm a=(-1,-1,\dots,-1)$, we have: 
\begin{equation}
	H(\widehat{\bm p}) = -\sum_{\ell=1}^k \widehat{p}_{\ell} \ln \widehat{p}_{\ell} \xrightarrow{\mathcal{D}} \mathcal N\big(H({\bm p}), \sigma^2_{n,\bm p}\big),
	\label{Eq:AsymptoticDistributionH}
\end{equation}	
where
\begin{multline}
	\sigma^2_{n,\bm p} = \frac{1}{n} \sum_{\ell=1}^k p_{\ell} (1-p_{\ell}) (\ln p_{\ell} + 1)^2 - \\
	\frac{2}{n} \sum_{{j}=1}^{k-1} \sum_{\ell=j+1}^{k} p_{j}p_{\ell} (\ln p_{j} + 1) (\ln p_{\ell} + 1).
	\label{Eq:AsymptoticVarianceH}
\end{multline}

Notice that the asymptotic mean coincides with Eq.~\eqref{eq:EntropyMultinomial}, and that computing the asymptotic variance does not pose any numerical difficulty.

Some studies and applications use a normalized entropy computed from Eq.~\eqref{eq:EntropyMultinomial} divided by $\ln k$. 
In this case, the asymptotic distribution is $\mathcal N\big(H({\bm p})/\ln k, \sigma^2_{n,\bm p}/(\ln k)^2\big)$.

\section{Experiments, numerical stability and accuracy}\label{Sec:Numerical}

Basharin's first-order approximations to the mean and variance, Eqs.~\eqref{Eq:FirstOrderMean} and~\eqref{Eq:FirstOrderVariance}, do not offer numerical difficulty, except in situations where there is at least one probability value near zero,  $p_\ell\approx 0$ for some $1 \leq \ell \leq k$ .
Such cases are easily handled as the limit $\lim_{p\to 0} p\ln p=0$ and pose no numerical challenges.
The same situation occurs with Second-order approximations, Eqs.~\eqref{Eq:ApproxMeanH} and~\eqref{Eq:ApproxVarH}.

Hutcheson's exact expressions, Eqs.~\eqref{Eq:ExactMeanH} and~\eqref{Eq:ExactVarH}, involve more sources of numerical instabilities, namely combinatorial numbers,  $\binom{a}{b}$ with very large  $a$ and $b$ values.
We implemented these exact expressions using computer algebra platforms 
(Mathematica, 
Yacas, and 
Maxima), as well as with high-precision specialized numerical functions in R (the \texttt{VeryLargeIntegers} package, that allows storing and operating with arbitrarily large integers).
None of these platforms returned useful values of either Eq.~\eqref{Eq:ExactMeanH} or Eq.~\eqref{Eq:ExactVarH} for practical situations.
Despite that, and for the sake of completeness, we report some of those results in the following.

Following \citet{AlmironetalJSS2010},
we compare a ``certified'' value $c$ with its ``approximation'' $x$ by computing the absolute value of the relative error and taking its decimal logarithm:
\begin{equation}
	\text{LRE}(x,c) = \begin{cases}
		-\log \frac{|x-c|}{|c|} & \text{if } c\neq0, \text{ and}\\
		-\log |x| & \text{otherwise}.
	\end{cases}
\end{equation}
This Log-Relative Error relates to the number of significant digits that are correctly computed, so we report its integer part $[\text{LRE}(x,c)]$.

In the following, the Basharin, Hutcheson, and Asymptotic approximations are compared with the exact expected value given in~\eqref{Eq:ExactMeanH}.

\begin{figure}
	\centering
	\includegraphics[width=\columnwidth]{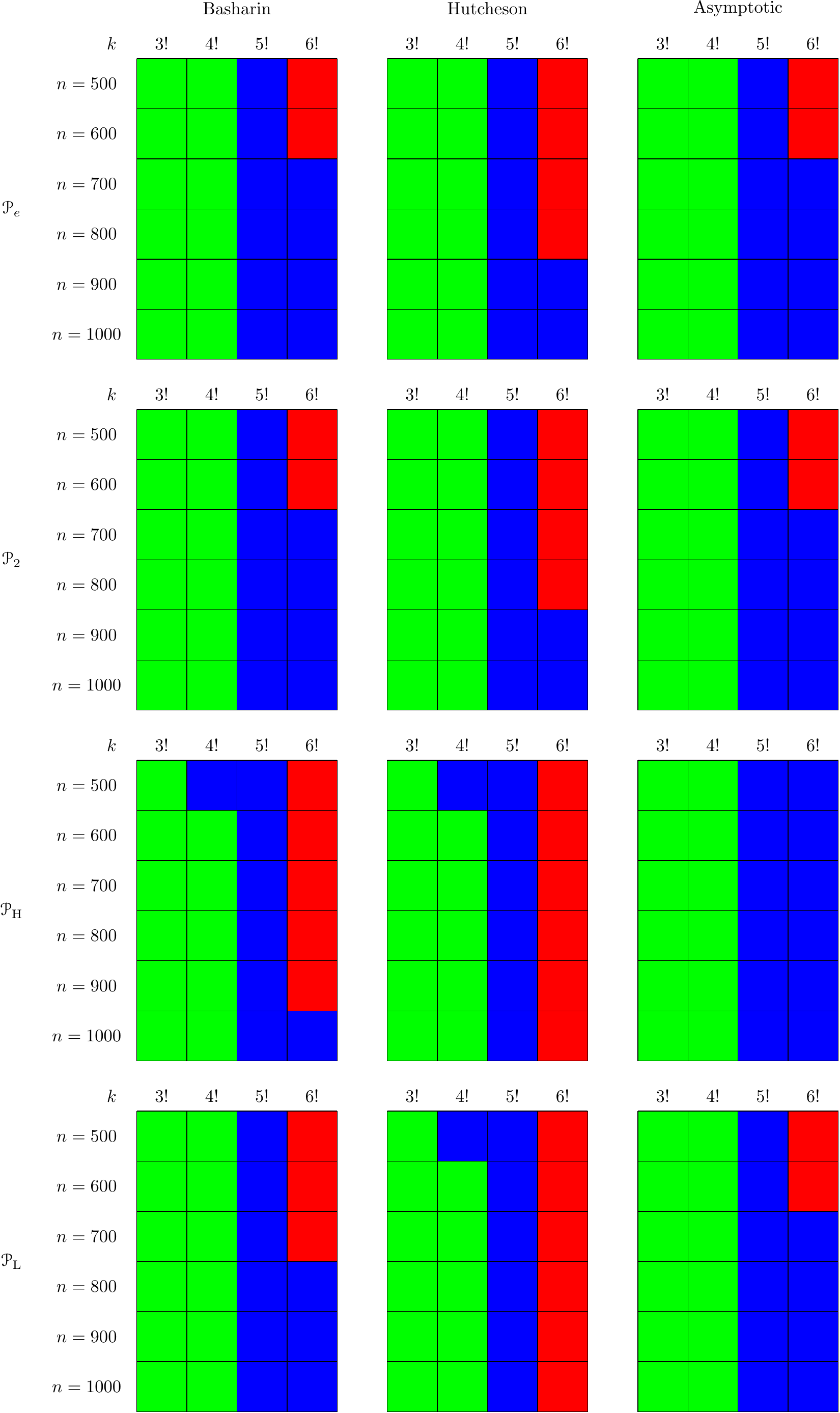}
	\caption{Log relative errors for Basharin, Hutcheson and asymptotic approximations, using different underlying distribution: $[\textrm{LRE}] = 2$ (green), $[\textrm{LRE}] = 1$ (blue), $[\textrm{LRE}] = 0$ (red).}
	\label{fig:lre}
\end{figure}

In this experiment, we applied four different types of underlying distributions: 
equiprobable $\mathscr{P}_e$, 
a perturbed equiprobable law $\mathscr{P}_2$, 
the half perturbed distribution $\mathscr{P}_{\textrm{H}}$, 
and the linear distribution $\mathscr{P}_{\textrm{L}}$. 
\begin{itemize}
	\item $\mathscr{P}_e$:  $p_\ell = \frac{1}{k}$,  $\ell =1, \dots, k$. 
	\item $\mathscr{P}_2$:  $p_\ell  = 1/k$,  $\ell =1, \dots, k-2$, $p_{k-1}=\frac{1}{k} +\varepsilon$ and $p_{k}=\frac{1}{k} -\varepsilon$,  $0<\varepsilon<1/k$.  
	\item $\mathscr{P}_{\textrm{H}}$:  $p_\ell  = 1/k -\varepsilon, \; \ell = 1, \dots, \frac{k}{2}$, and  $p_\ell  =1/k+\varepsilon, \; \ell = \frac{k}{2}, \dots, k $, $0<\varepsilon<1/k$
	\item $\mathscr{P}_{\textrm{L}}$:  $p_\ell  = \frac{\ell}{\sum_{j=1}^k j}$. 
\end{itemize}  
The results obtained are shown in Fig.~\ref{fig:lre}, where green, blue and red boxes correspond to $[\textrm{LRE}] = 2$, $1$, and $0$,  respectively.
It can be seen that the Asymptotic approximation has the best performance, followed by Basharin's approximation.

\begin{figure*}[hbt]
	\centering
	\includegraphics[width=.9\linewidth]{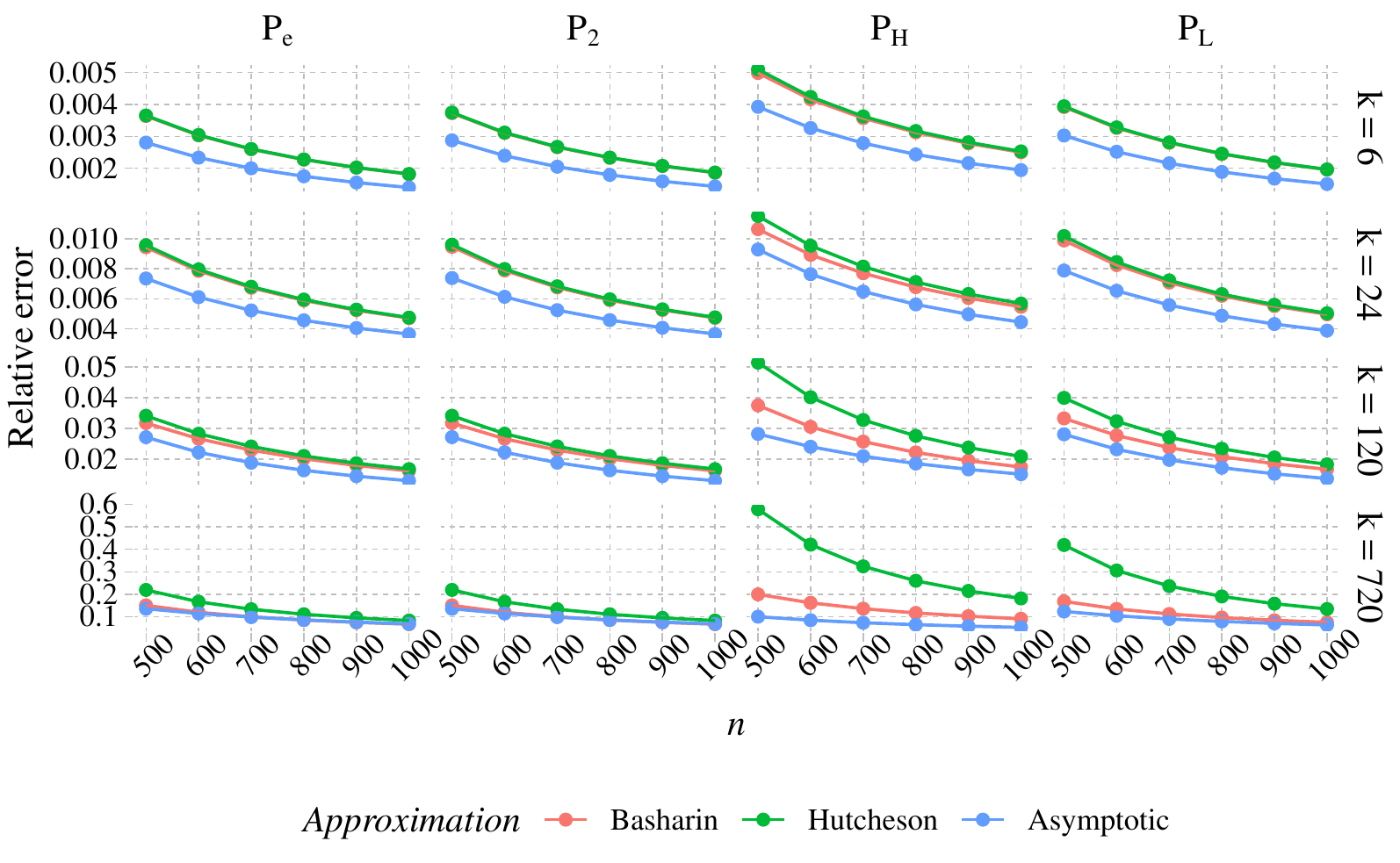}
	\caption{Relative errors for Basharin (pink), Hutcheson (green) and asymptotic (light blue) approximations, for a given number of ordinal patterns $k$ and different underlying distributions $\mathscr{P}_e$,   $\mathscr{P}_2$, $\mathscr{P}_{\textrm{H}}$ and  $\mathscr{P}_{\textrm{L}}$}
	\label{fig:relativeErrors}
\end{figure*}

The same scenarios and underlying distributions were used to compute the relative error 
\begin{equation}
	\text{RE}(x,c) = 
	\begin{cases}
		\frac{|x-c|}{|c|} & \text{if } c\neq0, \text{ and}\\
		|x| & \text{otherwise}.
	\end{cases}
\end{equation}
to compare the certified value $c=\operatorname{E}\big( H(\widehat{\bm p}) \big)$ given in~\eqref{Eq:ExactMeanH} against  the Basharin, Hutcheson and Asymptotic approximations.  The results obtained are shown in Fig.~\ref{fig:relativeErrors}, where it can be seen that for a previously selected  number of possible patterns \(k\) and any of the underlying distributions here applied,  the asymptotic estimate always provides the most accurate approximation (in terms of relative errors)

It is worth mentioning that we  computed  LRE and RE for $n\le 1000$ since, as already mentioned,  the formula for  the exact expression $\operatorname{E}\big( H(\widehat{\bm p}) \big)$ given in  Eq.~\eqref{Eq:ExactMeanH}, involves binomial coefficients that can only be computed with the aforementioned computational platforms when $n\le 1000$. 
Nevertheless, and despite the fact that these values of $n$ are not large enough to be considered ``asymptotic'', the asymptotic estimate  given in Eq.~\ref{Eq:AsymptoticDistributionH} is the most accurate approximation of  $\operatorname{E}\big( H(\widehat{\bm p}) \big)$  in terms of LRE and RE for any combination of $k$, $n$ and underlying distribution, even in those  situations where $k\approx n$.

As previously mentioned, Eqs.~\eqref{Eq:ExactMeanH} and~\eqref{Eq:ExactVarH} do not return usable values when implemented in either dependable numerical platforms or certified computer algebra systems.
For instance, Table~\ref{Tab:Usable} shows the maximum $n$ values for which  the exact mean under $\mathscr{P}_e$ and $\mathscr{P}_{\textrm{L}}$ can be obtained,  using Mathematica~\cite{Mathematica} in a computer \mbox{Mac OS X ARM} (\SI{64}{\bit}).
Notice that $\mathscr{P}_e$ poses harder numerical problems than $\mathscr{P}_{\textrm{L}}$ but, in any case, these numerical limitations make the exact values from Eqs.~\eqref{Eq:ExactMeanH} and~\eqref{Eq:ExactVarH} of little practical use; recall the rule-of-the-thumb $n\ge 100 k$.

\begin{table}[hbt]
	\caption{The largest $n$ such that Mathematica v.~13.0.0 returns a value for Eq.~\eqref{Eq:ExactMeanH} under two underlying distributions $\mathscr{P}$.}\label{Tab:Usable}
	\begin{tabular}{*5{c}}
		\toprule
		& \multicolumn{4}{c}{$k$}\\ \cmidrule{2-5}
		$\mathscr{P}$	& 6 & 24 & 120 & 720 \\ \midrule
		$\mathscr{P}_e$	& 400 & 228 & 151 & 110\\
		$\mathscr{P}_{\textrm{L}}$	& 571 & 283 & 174 & 122\\ \bottomrule
	\end{tabular}
\end{table}

\begin{figure*}[hbt]
	\centering
	\includegraphics[width=.9\linewidth]{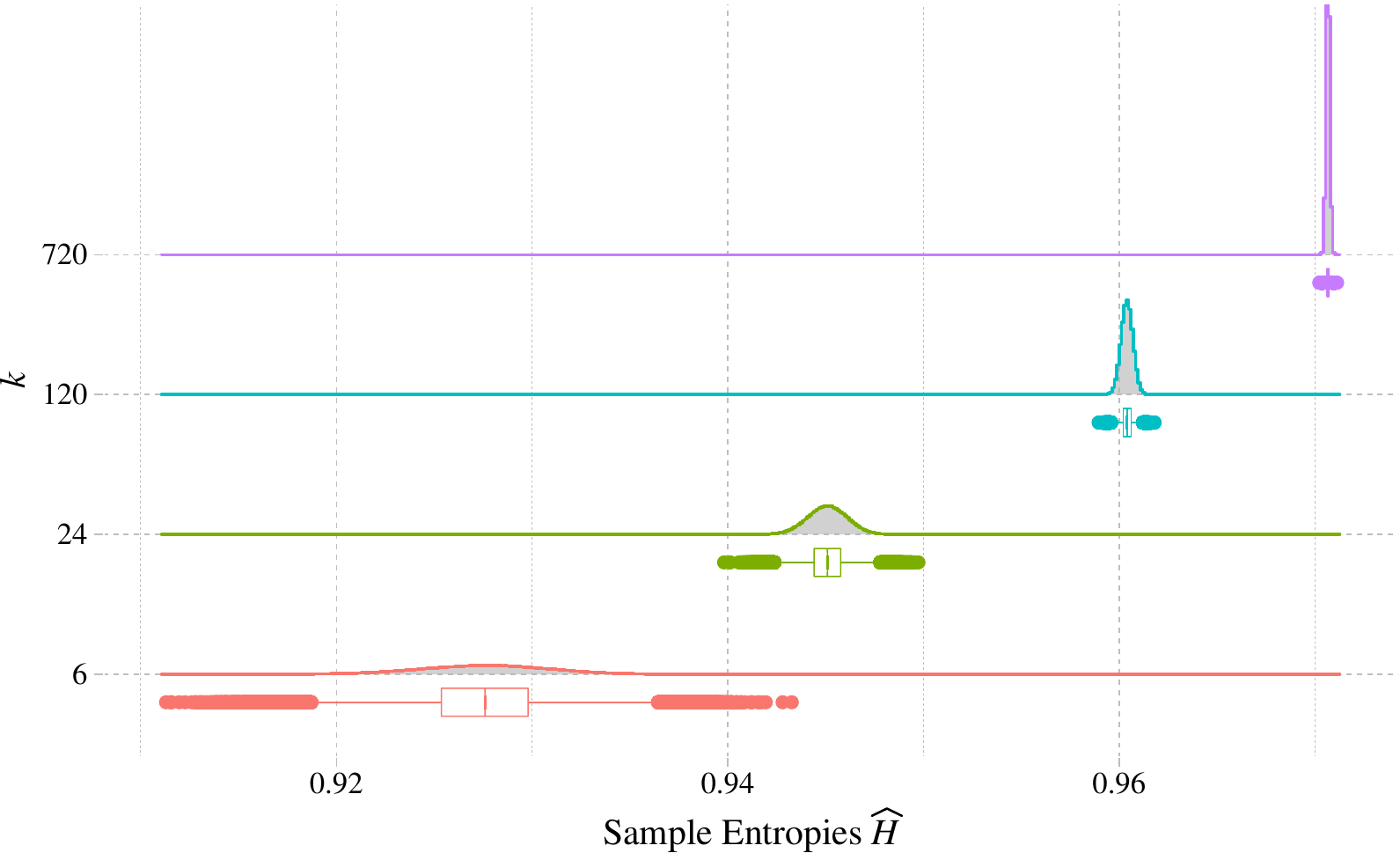}
	\caption{Histograms and boxplots of $10^6$ values of the entropy from the $\mathscr{P}_{\textrm{L}}$ model, $k\in\{6,24,120,720\}$, and series of length $n=10^3 k$.}
	\label{Fig:FourHistograms}
\end{figure*}

In order to study the behavior of expressions~\eqref{Eq:ExactMeanH},
\eqref{Eq:FirstOrderMean}, and
\eqref{Eq:ApproxMeanH} for long signals, in
the following
we consider as ``certified'' the value computed with independent simulations.
Firstly, we analyze the shape of the sample entropies from the $\mathscr{P}_{\text{L}}$ model.
For each value of $k\in\{6,24,120,720\}$, we sample $10^6$ independent values of $\widehat{H}$ from the $\mathscr{P}_{\text{L}}$ model and series of length $n=10^3 k$.
Fig.~\ref{Fig:FourHistograms} shows the histograms and boxplots of these data.

Table~\ref{Tab:Measures} shows the mean $\overline{\widehat{H}}$, median $q_{1/2}(\widehat{H})$, standard deviation $s({\widehat{H}})$, asymmetry $\gamma_1(\widehat{H})$ and excess kurtosis $\gamma_2(\widehat{H})$ of the observed entropy over $10^6$ independent samples from the $\mathscr{P}_{\text{L}}$ model, for four values of $k=D!$, along with approximate \SI{95}{\percent} confidence intervals. It can be observed that, as $k$ increases, so do the centrality measures of the samples (the mean $\overline{\widehat{H}}$ and the median $q_{1/2}(\widehat{H})$ coincide, suggesting no lack of symmetry).
The dispersion, measured by the standard deviation $s({\widehat{H}})$, reduces when $k$ increases.
The Asymmetry  value, $\gamma_1(\widehat{H})$ is slightly negative, and reduces with increasing $k$, and the excess kurtosis $\gamma_2(\widehat{H})$ is negligible in all cases.

\begin{table*}[hbt]
	\caption{Sample mean $\overline{\widehat{H}}$, median $q_{1/2}(\widehat{H})$, standard deviation $s({\widehat{H}})$, asymmetry $\gamma_1(\widehat{H})$ and excess kurtosis $\gamma_2(\widehat{H})$ of the observed entropy over $10^6$ independent samples from the $\mathscr{P}_{\text{L}}$ model, for four values of $k=D!$, along with approximate \SI{95}{\percent} confidence intervals.}\label{Tab:Measures}
	\begin{tabular}{c*5{c}}
		\toprule
		$k$ & 
		\multicolumn{1}{c}{$\overline{\widehat{H}}$} 
		& \multicolumn{1}{c}{$q_{1/2}(\widehat{H})$} 
		& \multicolumn{1}{c}{$s({\widehat{H}})$} 
		& \multicolumn{1}{c}{$\gamma_1(\widehat{H})$} 
		& \multicolumn{1}{c}{$\gamma_2(\widehat{H})$} \\  
		\cmidrule(lr){1-1} \cmidrule(lr){2-2} \cmidrule(lr){3-3} \cmidrule(lr){4-4} \cmidrule(lr){5-5} \cmidrule(lr){6-6}
		6	& 0.9276 & 0.9276 & 0.0033 & $-0.0693$ & 0.0033 \\  
		& \num{\pm6.44e-6} & \num{\pm8.09e-6} & \num{\pm 4.56e-6} & \num{\pm 4.80e-3} & \num{\pm 9.60e-3}\\ \cmidrule{2-6}
		24 	& 0.9451 & 0.9451 & 0.0010 & $-0.0296$ & 0.0004 \\
		& \num{\pm1.97e-6} & \num{\pm2.41e-6} & \num{\pm 1.40e-6} & \num{\pm 4.80e-3} & \num{\pm 9.60e-3}\\ \cmidrule{2-6}  
		120 & 0.9604 & 0.9604 & 0.0003 & $-0.0115$ & 0.0020 \\  
		& \num{\pm5.92e-7} & \num{\pm7.34e-7} & \num{\pm 4.19e-7} & \num{\pm 4.80e-3} & \num{\pm 9.60e-3}\\ \cmidrule{2-6}
		720 & 0.9707 & 0.9707 & 0.0001 & $-0.0033$ & $-0.0053$ \\ 
		& \num{\pm1.76e-7} & \num{\pm2.18e-7} & \num{\pm 1.24e-7} & \num{\pm 4.80e-3} & \num{\pm 9.60e-3}\\
		\bottomrule
	\end{tabular}
\end{table*}

\begin{figure*}[hbt]
	\centering
	\includegraphics[width=.9\linewidth]{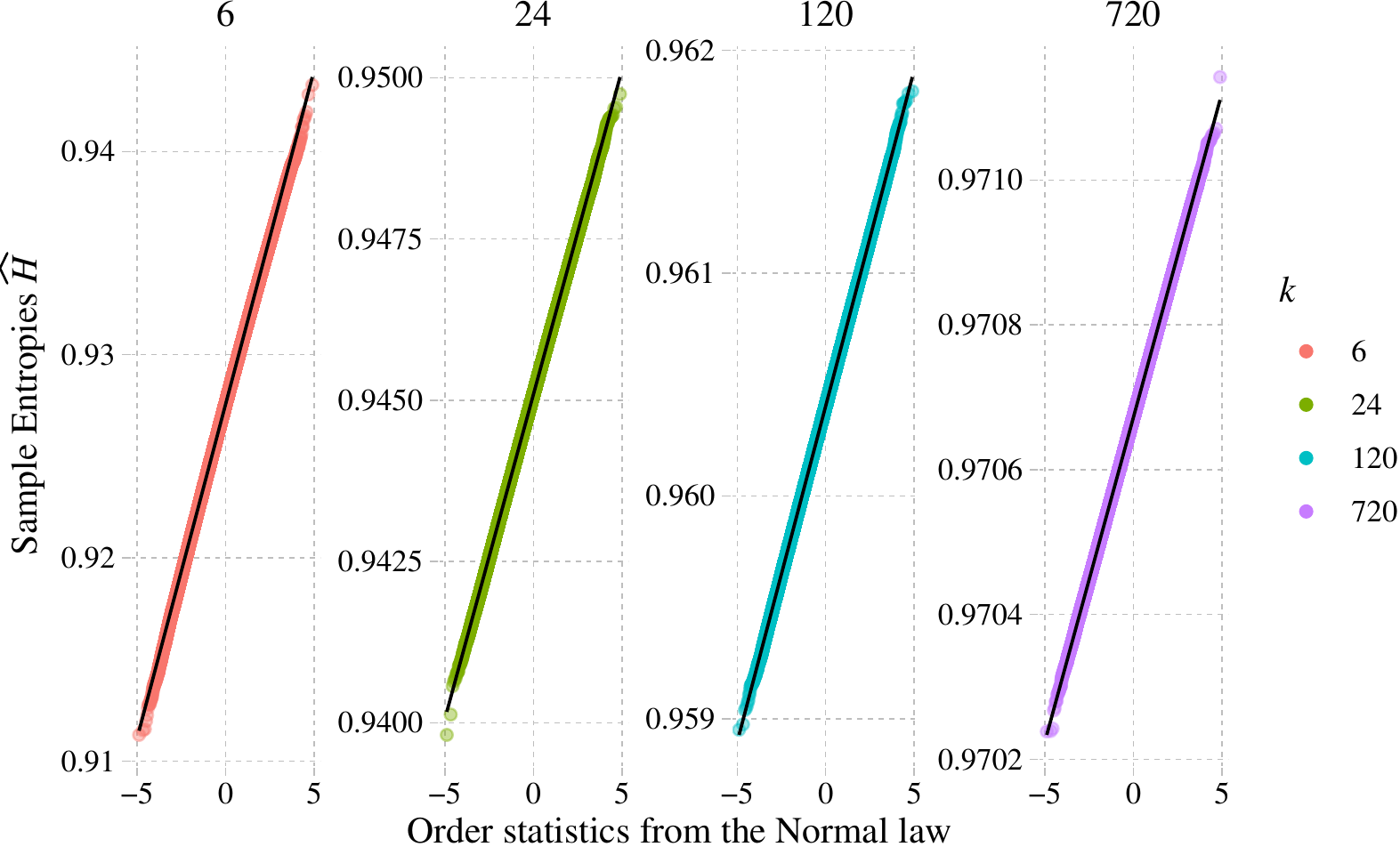}
	\caption{Quantile-quantile plots of the entropies for the Normal distribution.}
	\label{Fig:QQPlot}
\end{figure*}

Fig.~\ref{Fig:QQPlot} shows the quantile-quantile plots of the entropies for the Normal distribution. 
Notice that, although the histogram for $k=6$ does not reveal any clear deviation from a Normal density, the slight positive excess kurtosis explains the lack of fit of a few observations.
Apart from that, the observations lie very close to the straight lines.

\begin{figure*}[hbt]
	\centering
	\includegraphics[width=.9\linewidth]{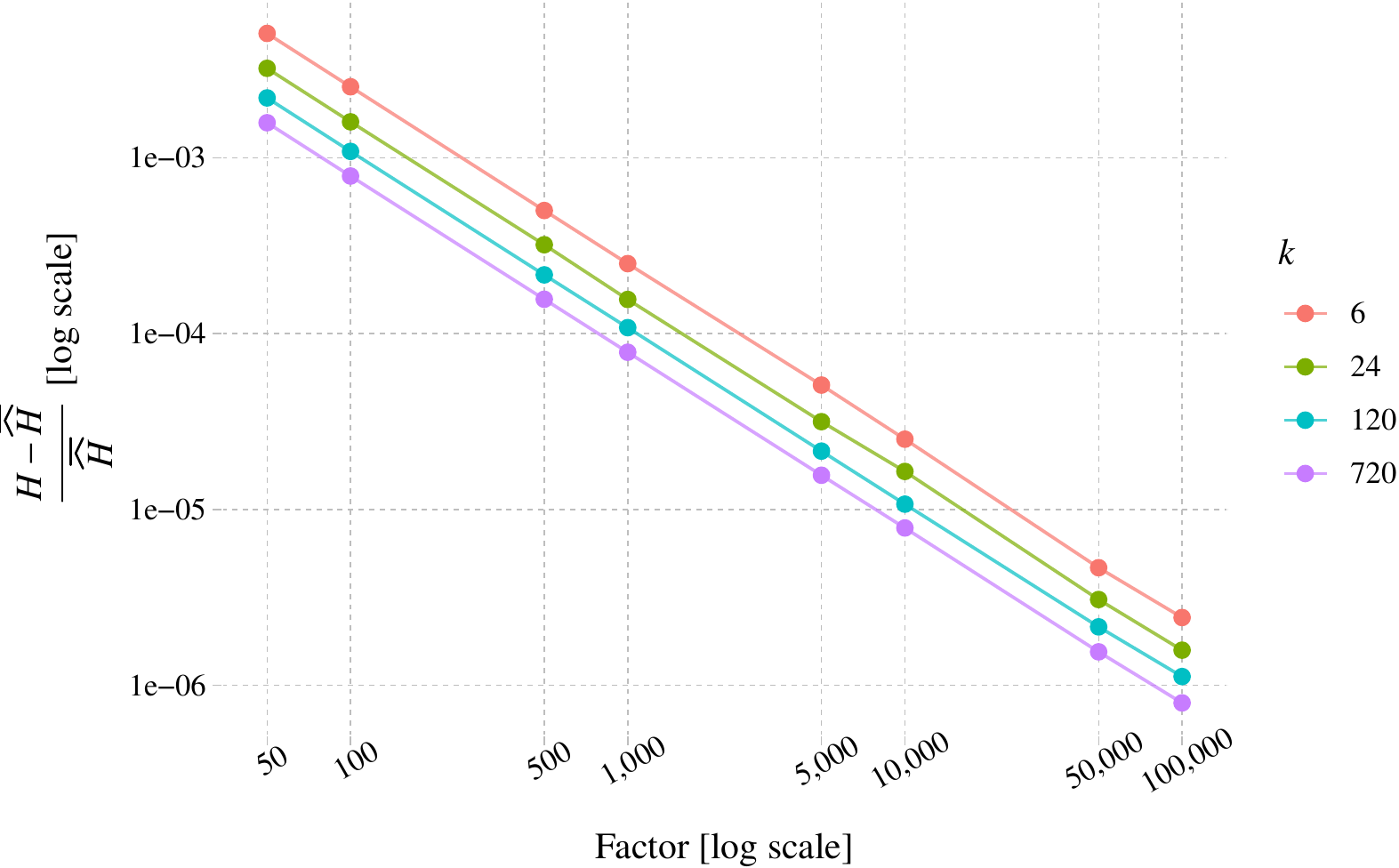}
	\caption{Relative error incurred by $H$ to the mean of $10^6$ independent samples $\overline{\widehat{H}}$, as a function of the number of possible patterns $k$ and of the factor that determines the number of observed patterns $n=\text{Factor }\times k$.}
	\label{Fig:Error_vs_Factor}
\end{figure*}

We now compare our ``certified'' values obtained by simulation with Hutcheson's exact values and third-order approximations.
Firstly, Fig.~\ref{Fig:Error_vs_Factor} shows, in log-log scales, the relationship between the relative error incurred by $H$ to the mean of $10^6$ independent samples $\overline{\widehat{H}}$, as a function of the number of possible patterns $k$ and of the factor that determines the number of observed patterns $n=\text{Factor}\times k$.
For a sample size fixed, the relative error decreases with the number of possible patterns.
When the number of possible patterns is fixed, the relative error decreases as the sample size increases.

\begin{figure*}[hbt]
	\centering
	\subcaptionbox{Case $k=6$ and $n=300$.\label{Fig:Hist6}}{\includegraphics[width=.48\linewidth]{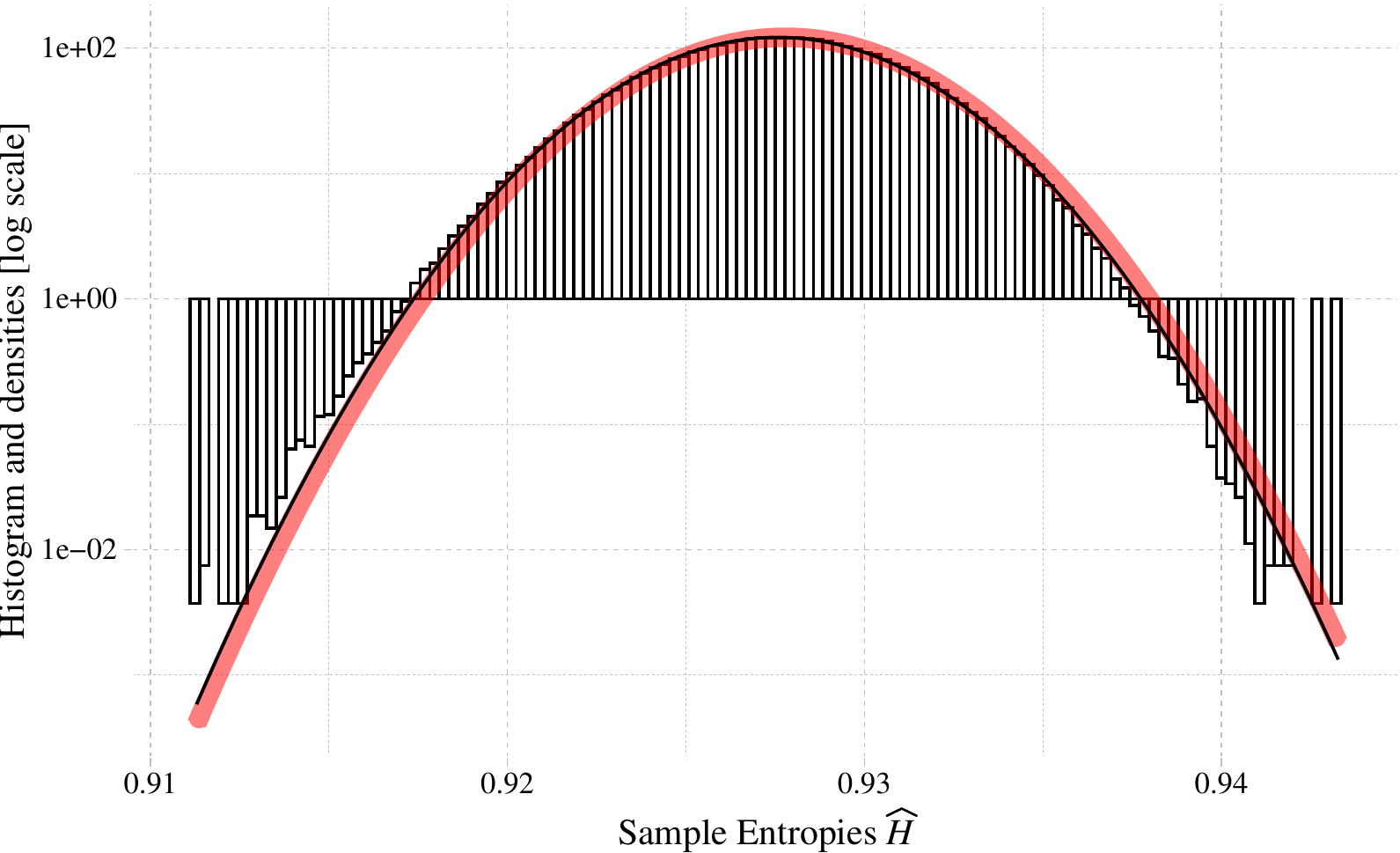}}
	\subcaptionbox{Case $k=720$ and $n=\num{72e6}$.\label{Fig:Hist720}}{\includegraphics[width=.48\linewidth]{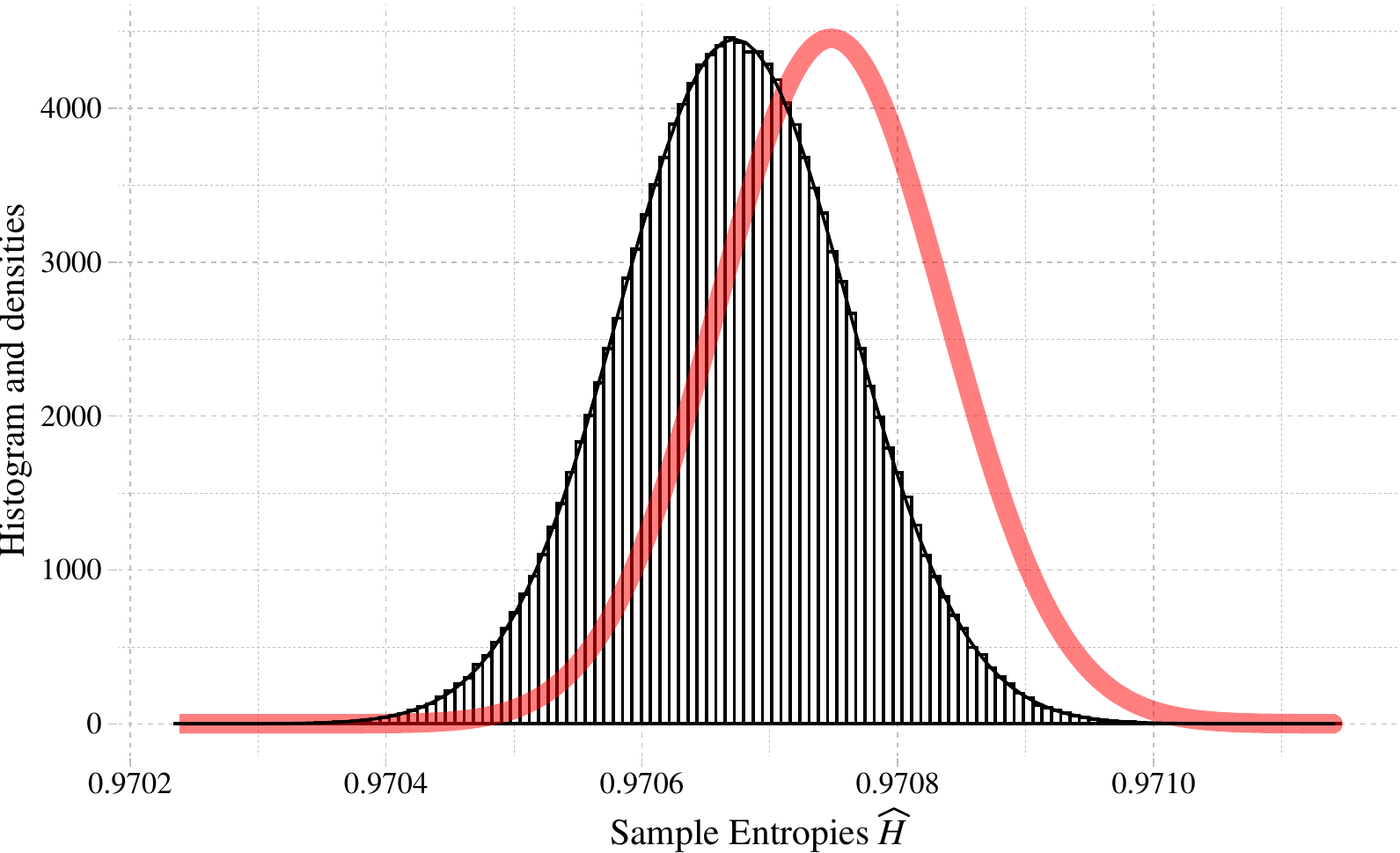}}	
	\caption{Histograms of $10^6$ samples of $\widehat{H}$ under $\mathscr{P}_{\text{L}}$, along with the empirical densities (thin black lines) and asymptotic models (thick red lines).}\label{Fig:Histograms}
\end{figure*}

The largest relative error is of order $5\times10^{-3}$; it corresponds to the case $k=6$ and $n=300$, for which $H\approx 0.92779$ and $\overline{\widehat{H}} \approx 0.92309$.
The smallest relative error occurs when $k=720$ and $n=\num{72e6}$.
In such a case, it amounts to approximately $\num{7.81987e-05}$.
Fig.~\ref{Fig:Histograms} shows, for each of the two extreme situations of minimum and maximum relative error, the histogram of $10^6$ observations, the underlying density function as a thin black line, and the asymptotic density distribution as a thick red line. We used a semilogarithmic scale.


It can be observed that the asymptotic standard deviation is acceptable for describing the data dispersion and the asymptotic mean deviates from the observed mean, as   Fig.~\ref{Fig:Error_vs_Factor} shows.
Such deviation is slight in the case shown in Fig.~\ref{Fig:Hist6}, and much more noticeable in Fig.~\ref{Fig:Hist720}.
With this, we conclude that the relative error is an unreliable measure of the quality of a model.

Figs.~\ref{Fig:Error_vs_Factor} and~\ref{Fig:Histograms} suggest, thus, that there is room for an improved asymptotic model for the sample entropy $\widehat{H}$.
Our proposal for a corrected model stems from fusing the information we have about the asymptotic distribution of $\widehat{H}$, namely Eqs.~\eqref{Eq:AsymptoticDistributionH} and~\eqref{Eq:AsymptoticVarianceH} and the third-order corrected expected value of $\widehat{H}$ given in Eq.~\eqref{Eq:ApproxMeanH}.
With this, given the sequence $(\pi_1,\pi_2,\dots,\pi_n)$ of symbols obtained from words of size $D$, 
first compute its histogram of proportions $\widehat{\bm p} = (\widehat p_1, \widehat p_2,\dots, \widehat p_{k})$, 
where $k=D!$ and $\widehat p_\ell=\#\{j:\pi_j=\pi^{(\ell)}, 1\leq j\leq n\}/n$, and 
then assume that $\widehat{H}=-\sum_{\ell=1}^{k}\widehat p_\ell \ln \widehat p_\ell$ is an outcome from a random variable that has Normal distribution with mean
\begin{multline}
	\mu_{n,{\bm p}}  = -\sum_{\ell=1}^{k} p_\ell \ln  p_\ell
	- \frac{k-1}{2n} + \frac{1-\sum_{\ell=1}^k  p_\ell^{-1}}{12 n^2} + \\
	\frac{\sum_{\ell=1}^k ( p_\ell^{-1}- p_\ell^{-2})}{12 n^3}	,
	\label{Eq:ProposedMeanH}
\end{multline}
and variance given by 
\begin{multline}
	\sigma^2_{n,{\bm p}} = \frac{1}{n} \sum_{\ell=1}^k  p_{\ell} (1- p_{\ell}) (\ln  p_{\ell} + 1)^2 - \\
	\frac{2}{n} \sum_{{j}=1}^{k-1} \sum_{\ell=j+1}^{k}  p_{j} p_{\ell} (\ln  p_{j} + 1) (\ln  p_{\ell} + 1).
	\label{Eq:ProposedVarianceH}
\end{multline}

\begin{figure*}[hbt]
	\subcaptionbox{Case $k=6$ and $n=\num{300}$.\label{Fig:HistogramCorrectedDensityk6}}{\includegraphics[width=.48\linewidth]{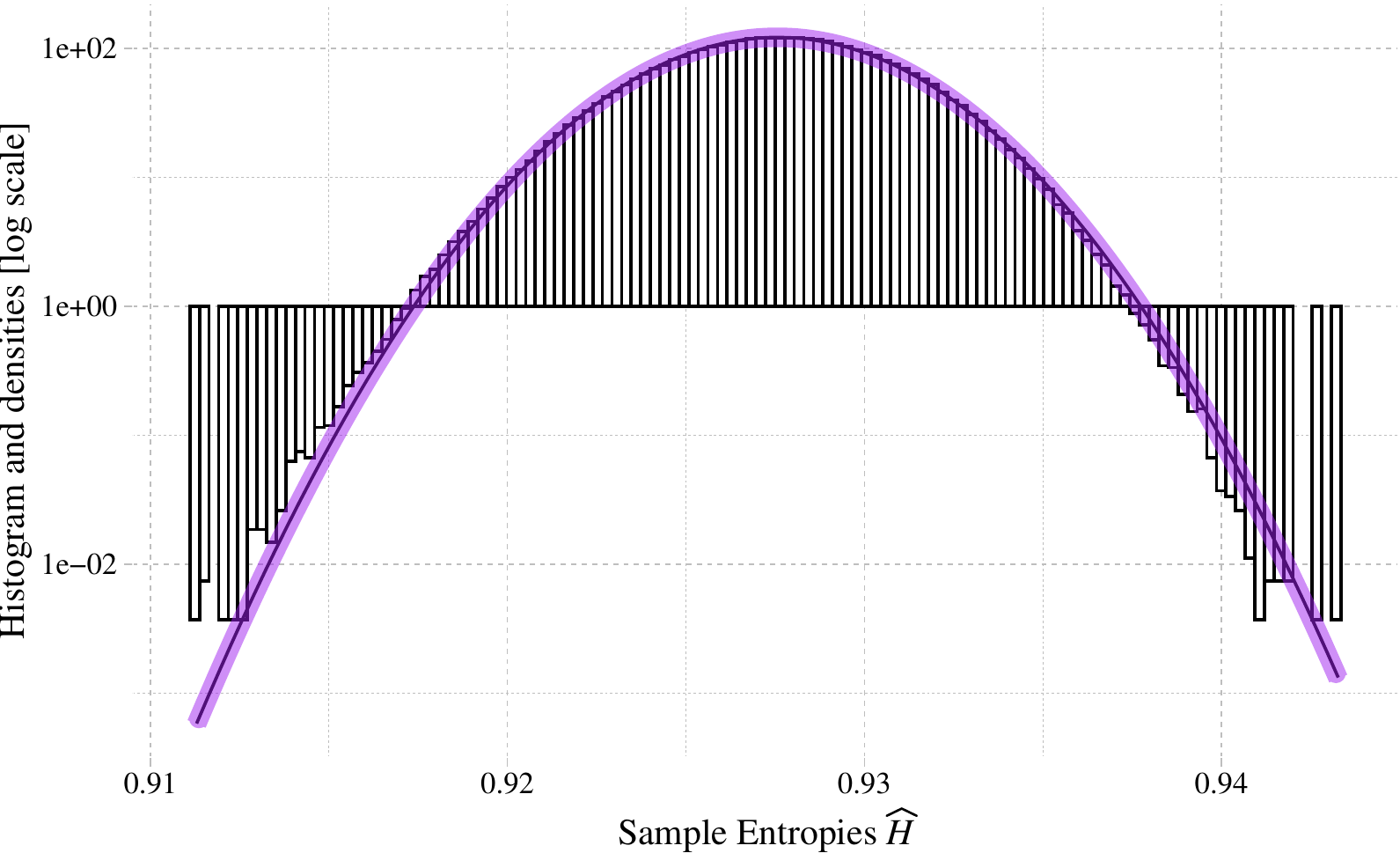}}
	\subcaptionbox{Case $k=720$ and $n=\num{72e6}$.\label{Fig:HistogramCorrectedDensityk720}}{\includegraphics[width=.48\linewidth]{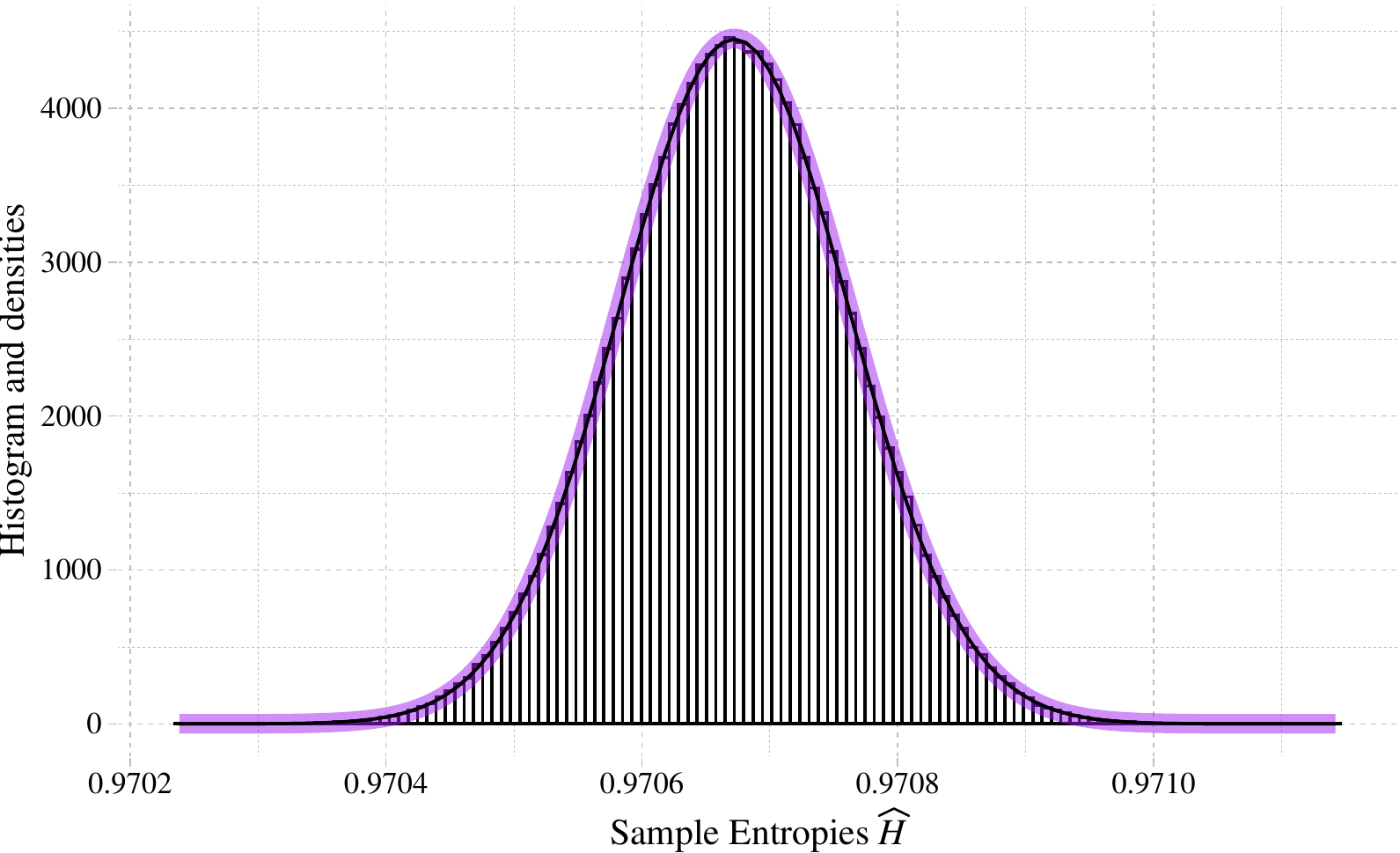}}
	\caption{Histograms of $10^6$ samples of $\widehat{H}$ under $\mathscr{P}_{\text{L}}$, along with the sample density (thin black line) and the corrected model (purple thick line).}
	\label{Fig:CorrectedModels}
\end{figure*}

Fig.~\ref{Fig:CorrectedModels} shows the histograms of $10^6$ samples of $\widehat{H}$ in two situations, namely $k=6$ and $n=\num{300}$ (Fig.~\ref{Fig:HistogramCorrectedDensityk6}), and 
$k=720$ and $n=\num{72e6}$ (Fig.~\ref{Fig:HistogramCorrectedDensityk720}).
In both cases, the samples were produced under $\mathscr{P}_{\text{L}}$.
The empirical densities are shown in thin black lines,
while the corrected densities are shown in thick purple lines.
The corrected model, i.e, the Normal distribution with mean $\mu_{n, \widehat{\bm p}}$ and variance $\sigma^2_{n, \widehat{\bm p}}$, provides an excellent description of the observed data.

\section{Hypothesis tests and applications}\label{Sec:Hypothesis}

Let \(\bm x = (x_1, x_2, \dots, x_{n_{\bm{x}}})\) and \(\bm y = (y_1, y_2, \dots, y_{n_{\bm{y}}})\) be two independent time series of length
\(n_{\bm{x}} = T_{\bm{x}}+D_{\bm{x}}-1\) and \(n_{\bm{y}} = T_{\bm{y}}+D_{\bm{y}}-1\), respectively. 
Let us also assume that these time series were generated by the stochastic (and maybe unknown) models $\mathcal{M}_{\bm{x}}$ and $\mathcal{M}_{\bm{y}}$.
We are interested in testing if these time series are statistically different.
We choose to compare the entropy of their ordinal patterns among the many possible ways of performing such an assessment.

In this sense, we compute the series of symbols \(\pi^{\bm{x}}_1, \pi^{\bm{x}}_2, \dots, \pi^{\bm{x}}_{D_{\bm{x}}!}\) and \(\pi^{\bm{y}}_1, \pi^{\bm{y}}_2, \dots, \pi^{\bm{y}}_{D_{\bm{y}}!}\), and then we find the histograms \(\widehat{\bm p}_{\bm{x}}\) and \(\widehat{\bm p}_{\bm{y}}\).
From Section~\ref{Sec:Asymptotic}, we know that the Shannon's Entropies $H(\widehat{\bm p}_{\bm{x}})$ and $H(\widehat{\bm p}_{\bm{y}})$, are random variables with asymptotic distributions 
\( \mathcal N \big(\mu_{n_{\bm x}, {\bm p}_{\bm x}}, \sigma^2_{n_{\bm{x}},{\bm p}_{\bm{x}}}\big)\) 
and 
\( \mathcal N\big(\mu_{n_{\bm y}, {\bm p}_{\bm y}}, \sigma^2_{n_{\bm{y}}, {\bm p}_{\bm{y}}}\big),\)  respectively, 
with means and variances given by Eqs.~\eqref{Eq:ProposedMeanH} and~\eqref{Eq:ProposedVarianceH}.

If \(\bm x\) and \(\bm y\) have the same underlying dynamics, then 
\(H({\bm p}_{\bm{x}}) =H({\bm p}_{\bm{y}})\), and 
we expect to observe \(H(\widehat{\bm p}_{\bm{x}}) \approx H(\widehat{\bm p}_{\bm{y}})\).
Our test will then verify the following hypothesis:
\begin{align*}
	\mathcal H_0 &: H({\bm p}_{\bm{x}})-H({\bm p}_{\bm{y}})=0,\\
	\intertext{versus the alternative}  
	\mathcal H_1 &: H(\bm p_{\bm{x}})-H(\bm p_{\bm{y}}) \neq 0.
\end{align*}
It is, therefore, a bilateral test.

Our test statistic is 
\begin{equation}
	W = H(\widehat{\bm p}_{\bm{x}}) - H(\widehat{\bm p}_{\bm{y}}).
	\label{Eq:TestStatisticW}
\end{equation}
Using~\eqref{Eq:ProposedMeanH} and~\eqref{Eq:ProposedVarianceH} and the assumption that the time series $\bm x$ and $\bm y$ are independent, it is straightforward that $W \xrightarrow{\mathcal{D}} \mathcal N\big(\mu_W, \sigma^2_W\big)$, where 
$\mu_W = \mu_{n_{\bm x}, {\bm p}_{\bm x}} - \mu_{n_{\bm y}, {\bm p}_{\bm y}}$, 
and 
$\sigma^2_W=\sigma^2_{n_{\bm{x}},{\bm p}_{\bm{x}}} + \sigma^2_{n_{\bm{y}}, {\bm p}_{\bm{y}}}$. 
Thus, for any observed $\eta >0$, holds that
\begin{multline}
	\Pr\big( |H(\widehat{\bm p}_{\bm{x}}) - H(\widehat{\bm p}_{\bm{y}})| \leq \eta \mid \mathcal H_0\big)  \approx  \Pr\big(|W| \leq \eta \mid \mathcal H_0\big) \\
	=   2\Phi\Big(\frac{\eta-\mu_W}{\sigma_W}\Big)-1,
\end{multline}
where  $\Phi$ is the standard Gaussian cumulative distribution function.

Under $\mathcal H_0$, the test statistic $W$ is asymptotically distributed as $\mathcal{N}\big(0, \sigma^2_W\big)$ and therefore, the $p$-value of the observed test statistic is approximately $2\big(1-\Phi(\epsilon)\big)$, where 
\begin{align}
\epsilon = \frac{H({\widehat{\bm p}_{\bm x}})- H({\widehat{\bm p}_{\bm y}})}{\widehat{\sigma}_W}, \label{Eq:Epsilon}\\
\intertext{and}
\widehat{\sigma}_W=\sqrt{\sigma^2_{n_{\bm{x}},{\widehat{\bm p}}_{\bm{x}}} + \sigma^2_{n_{\bm{y}}, {\widehat{\bm p}}_{\bm{y}}}}.
\end{align}

Unilateral tests and other specific hypotheses can be easily obtained using that $H(\widehat{\bm p}_{\bm{x}}) \sim \mathcal N \big(\mu_{n_{\bm x}, \widehat{\bm p}_{\bm x}}, \sigma^2_{n_{\bm{x}},\bm p_{\bm{x}}}\big)$ 
and that $H(\widehat{\bm p}_{\bm{y}}) \sim \mathcal N\big(\mu_{n_{\bm y}, \widehat{\bm p}_{\bm y}}, \sigma^2_{n_{\bm{y}},\bm p_{\bm{y}}}\big)$.
In particular, for the unilateral tests:
\begin{enumerate}
	\item $\mathcal H_0 : H({\bm p}_{\bm{x}})-H({\bm p}_{\bm{y}}) \geq 0 \text{ vs. }  
	\mathcal H_1 : H(\bm p_{\bm{x}}) - H(\bm p_{\bm{y}}) < 0$,
	\item $\mathcal H_0 : H({\bm p}_{\bm{x}})-H({\bm p}_{\bm{y}}) \leq 0 \text{ vs. }  
	\mathcal H_1 : H(\bm p_{\bm{x}}) - H(\bm p_{\bm{y}}) > 0$,
\end{enumerate}
the $p$-values of the observed test statistics are approximately $\Phi(\epsilon)$ and $1-\Phi(\epsilon)$, respectively, with $\epsilon$ defined in~\eqref{Eq:Epsilon}.
These approximations follow the approach discussed by \citet[][Sec.~3.6.1]{EssentialStatisticalInferenceTheoryMethods}.

\section{Climate Data}\label{Sec:Climate}

Our goal in this analysis is to evaluate sequences that come from:
(i)~the same underlying process and,
(ii)~different dynamics.
For this, we analyzed series of maximum daily temperatures measured in three  stations around the world: Dublin Phoenix Park (Ireland), 
Edinburgh Royal Botanic Garden (Scotland), and 
Miami International Airport (United States of America).
We obtained these datasets from the Climate Data Online website, supported by the National Oceanic and Atmospheric Administration (NOAA) at \url{https://www.ncei.noaa.gov/cdo-web/}.
They are part of the
GHCN (Global Historical Climatology Network)-Daily data set, the world's most extensive collection of daily climatology measurements, that  functions as the official archive for daily data from the Global Climate Observing
System (GCOS) Surface Network (GSN).
This data set contains observations of a variety of meteorological elements, including maximum and minimum temperature, 
at more than \num{100000} stations distributed across all continents.

\begin{figure*}
	\centering
	\includegraphics[width=\linewidth]{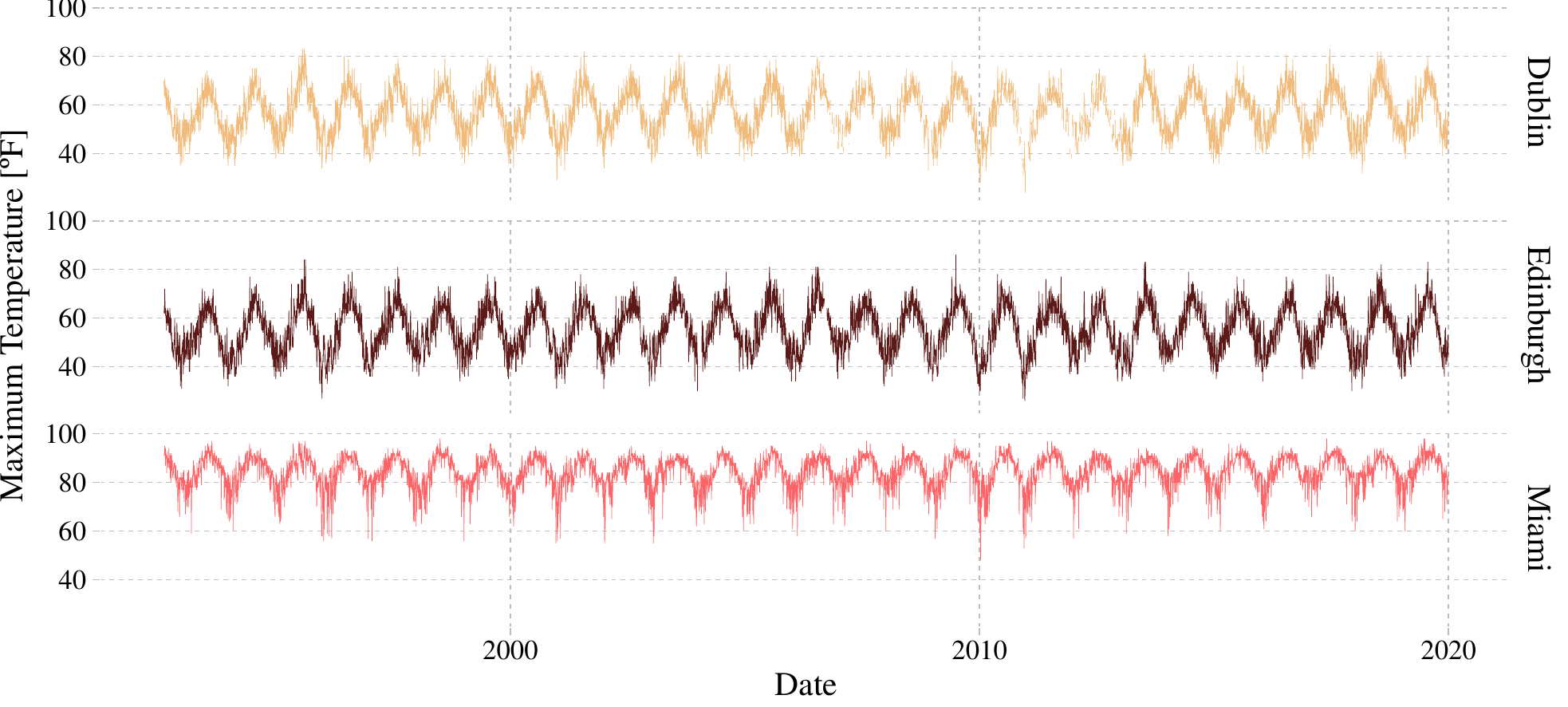}
	\includegraphics[width=.45\linewidth]{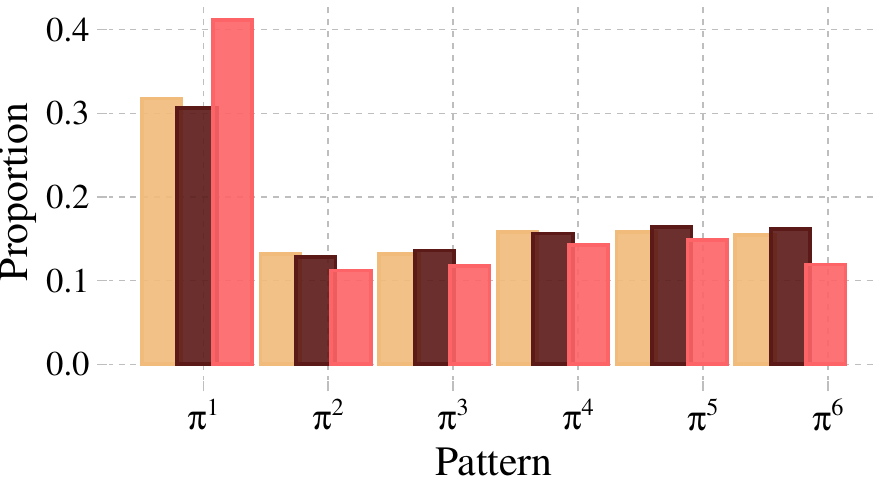}
	\includegraphics[width=.45\linewidth]{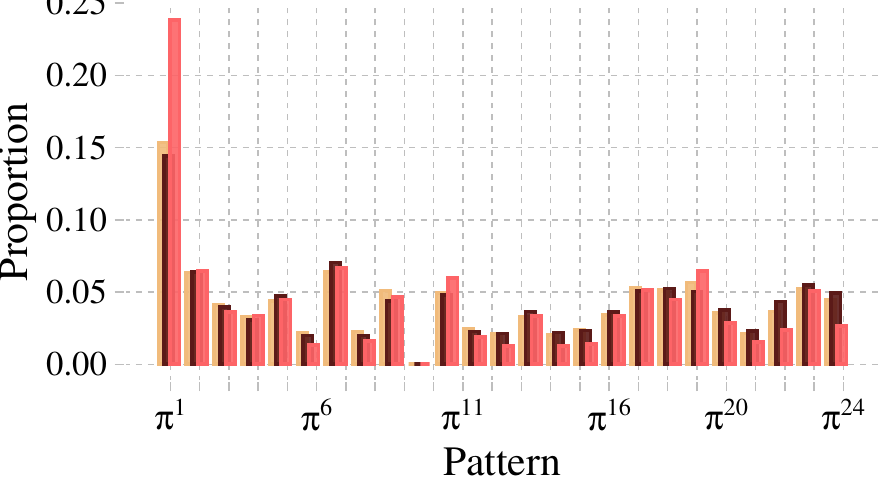}
	\caption{Maximum daily temperatures in Dublin, Edinburgh and Miami, from 8 August 1992 until 30 December 2019, along with the histograms of $k=6$ (left) and $k=24$ (right) patterns.}
	\label{fig:climate_data}
\end{figure*}

We used daily data from 
8 August 1992 until 30 December 2019.
These daily observations are shown in Fig.~\ref{fig:climate_data}, in Fahrenheit degrees.

Dublin and Edinburgh have similar temperate oceanic climates, with few temperature differences. 
However, extreme temperatures might be a little below zero Celsius degrees (\SI{32}{\degree \farad}) a few days of the winter and make it into the mid-seventies for some summer days. 
On the other hand, Miami has a tropical monsoon climate, with an average temperature of \SI{77}{\degree\farad} (\SI{25}{\degreeCelsius})
and extreme temperatures of \SI{50}{\degree \farad} in winter and \SI{91.4}{\degree \farad} in summer (\SI{10}{\degreeCelsius} and \SI{33}{\degreeCelsius}, respectively).
Fig.~\ref{fig:climate_data} shows that, beyond these differences in the marginal properties, the series from Dublin and Edinburgh seem to follow a simple sine-cosine plus noise pattern closely. In contrast, the dynamics underlying the Miami data seem more complex: there is less variation except during January and February, where the maximum temperatures exhibit large deviations.

Fig.~\ref{fig:climate_data} also shows the histograms of $k=6$ (left) and $k=24$ (right) patterns.
Dublin and Edinburgh have similar histograms, while Miami's data has a predominant pattern.
This configuration also appears when analyzing $k=120$ and $k=720$ patterns, but the visualization is too busy.
Notice that it is impossible to make a visual comparison of histograms obtained with different embedding dimensions, although our test statistics allow for such an operation.

We applied the test introduced in Section~\ref{Sec:Hypothesis} to contrast the null hypothesis between each possible pair of locations considering different values of the embedding dimension $D\in\{3,4,5,6\}$ and, thus, of the number of possible patterns $k\in\{6,24,120,720\}$.
In this case, all the time series have the same length $n=\num{10000}$. 
The corresponding $p$-values are shown in Table~\ref{Tab:pvalues}, from which we can see that $H_0$ is not rejected only for Dublin versus Edinburgh except for $k$ taking larger values for Dublin than for Edinburgh, or $k=6$ and $k=24$ being the same value for both zones.
These conclusions hold even after applying a conservative Bonferroni corection\cite{EffectSizeConfidenceIntervalandStatisticalSignificanceaPracticalGuideforBiologists}.

\begin{table*}[hbt]
	\caption{$p$-values of the hypothesis test applied to maximum temperature data.}\label{Tab:pvalues}
	
	Dublin vs.\ Edinburgh
	
	\begin{tabular}{c*4{c}}
		\toprule
		\diagbox{Dublin}{Edinburgh}
		& \multicolumn{1}{c}{$k=6$} 
		& \multicolumn{1}{c}{$k=24$} 
		& \multicolumn{1}{c}{$k=120$} 
		& \multicolumn{1}{c}{$k=720$} 
		\\  
		\cmidrule(lr){2-2} \cmidrule(lr){3-3} \cmidrule(lr){4-4} \cmidrule(lr){5-5} 
		$k=6$	& 0.0005 & 0.3294 & 0.6766 & 0.0830 \\ 
		$k=24$ 	& \num{4.60e-05} & 0.0457 & 0.4694 & 0.6354 \\ 
		$k=120$ & \num{1.48e-06} & 0.0047 & 0.1124 & 0.6639  \\ 
		$k=720$ & \num{4.98e-09} & 0.0001 & 0.0106 & 0.1594 \\
		\bottomrule
	\end{tabular}
	
	\medskip
	
	Dublin vs.\ Miami
	
	\begin{tabular}{c*4{c}}
		\toprule
		\diagbox{Dublin}{Miami}
		& \multicolumn{1}{c}{$k=6$} 
		& \multicolumn{1}{c}{$k=24$} 
		& \multicolumn{1}{c}{$k=120$} 
		& \multicolumn{1}{c}{$k=720$} 
		\\  
		\cmidrule(lr){2-2} \cmidrule(lr){3-3} \cmidrule(lr){4-4} \cmidrule(lr){5-5} 
		$k=6$	& \num{8.78e-08} & \num{1.68e-11} & \num{3.33e-15} & 0 \\  
		$k=24$ 	& \num{4.85e-03} & \num{7.71e-06} & \num{1.47e-08} & \num{5.98e-12} \\ 
		$k=120$ & \num{1.48e-01} & \num{1.54e-03} & \num{1.26e-05} & \num{2.89e-08}  \\  
		$k=720$ & \num{8.66e-01} & \num{4.87e-05} & \num{1.37e-03} & \num{1.21e-05} \\
		\bottomrule
	\end{tabular}
	
	\medskip
	
	Edinburgh vs.\ Miami 
	
	\begin{tabular}{c*4{c}}
		\toprule
		\diagbox{Edinburgh}{Miami}
		& \multicolumn{1}{c}{$k=6$} 
		& \multicolumn{1}{c}{$k=24$} 
		& \multicolumn{1}{c}{$k=120$} 
		& \multicolumn{1}{c}{$k=720$} 
		\\  
		\cmidrule(lr){2-2} \cmidrule(lr){3-3} \cmidrule(lr){4-4} \cmidrule(lr){5-5} 
		$k=6$	& 0 & 0 & 0 & 0 \\  
		$k=24$ 	&\num{2.86e-07}  & \num{9.05e-11} & \num{4.33e-14} & 0 \\ 
		$k=120$ & \num{7.86e-04} & \num{1.20e-06} & \num{2.49e-09} & \num{1.25e-12}  \\ 
		$k=720$ & \num{5.83e-02} & \num{4.57e-04} & \num{3.30e-06} & \num{7.13e-09} \\
		\bottomrule
	\end{tabular}
\end{table*}

\begin{figure*}[hbt]
	\centering
	\includegraphics[width=.245\linewidth]{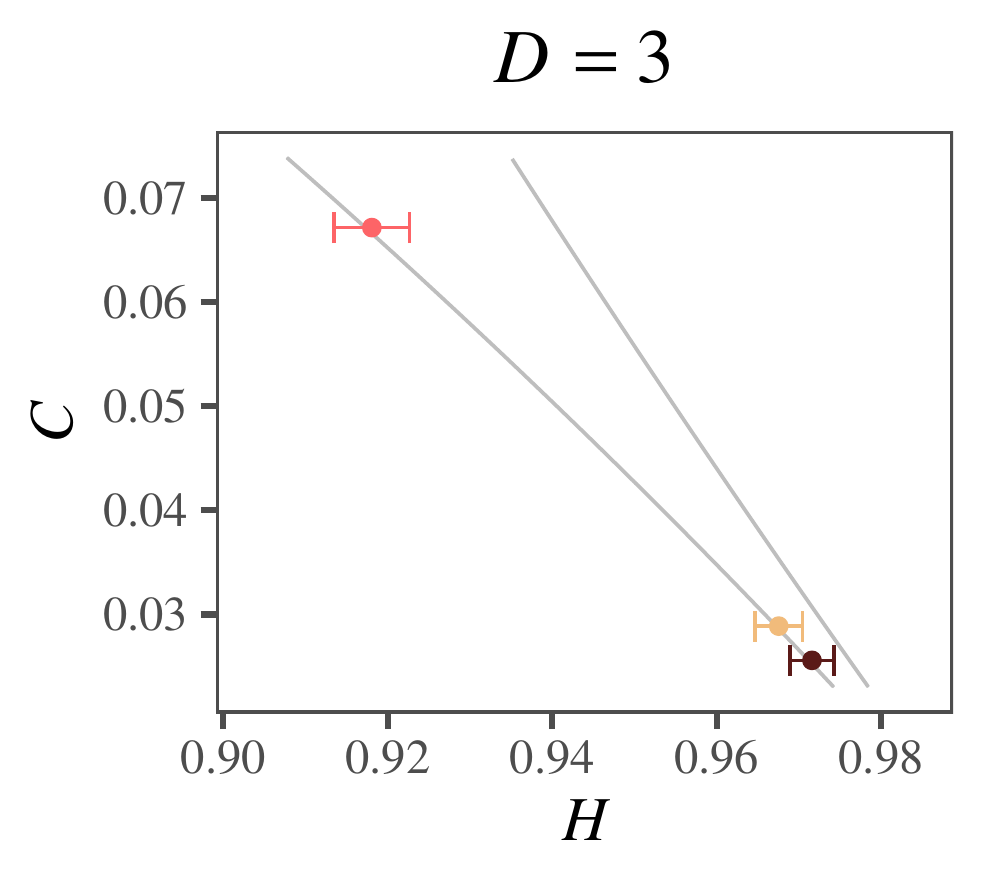}
	\includegraphics[width=.245\linewidth]{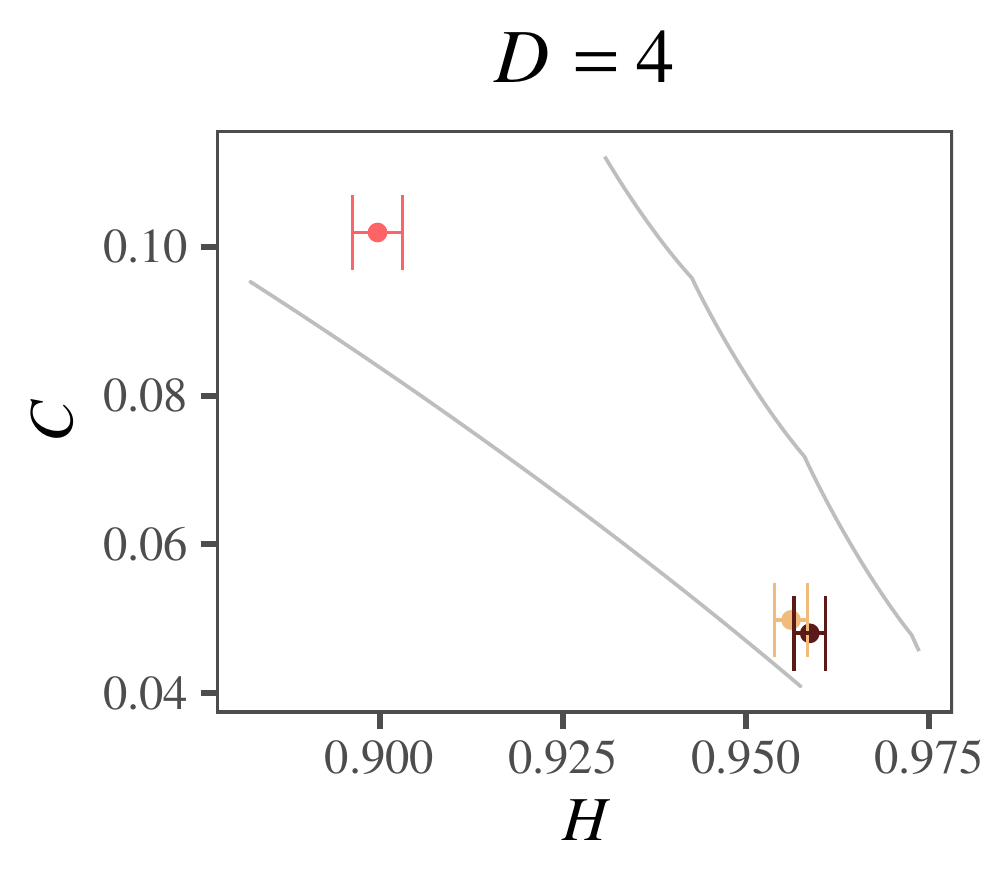}
	\includegraphics[width=.245\linewidth]{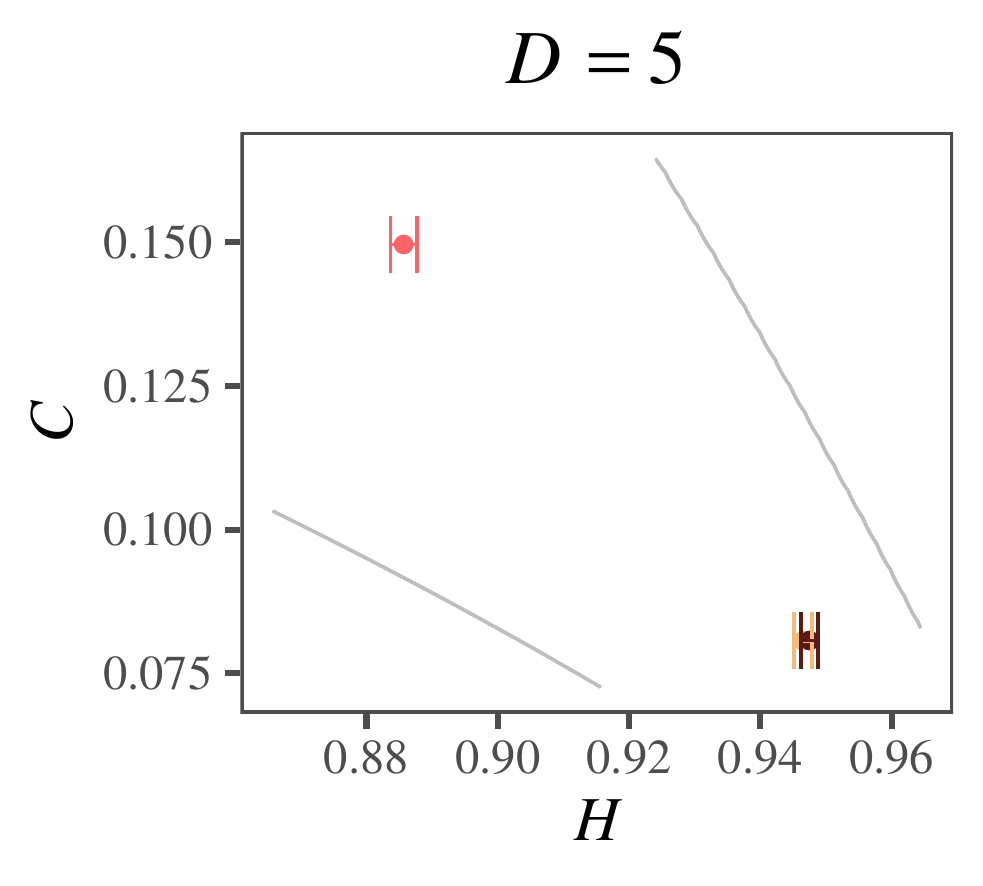}
	\includegraphics[width=.245\linewidth]{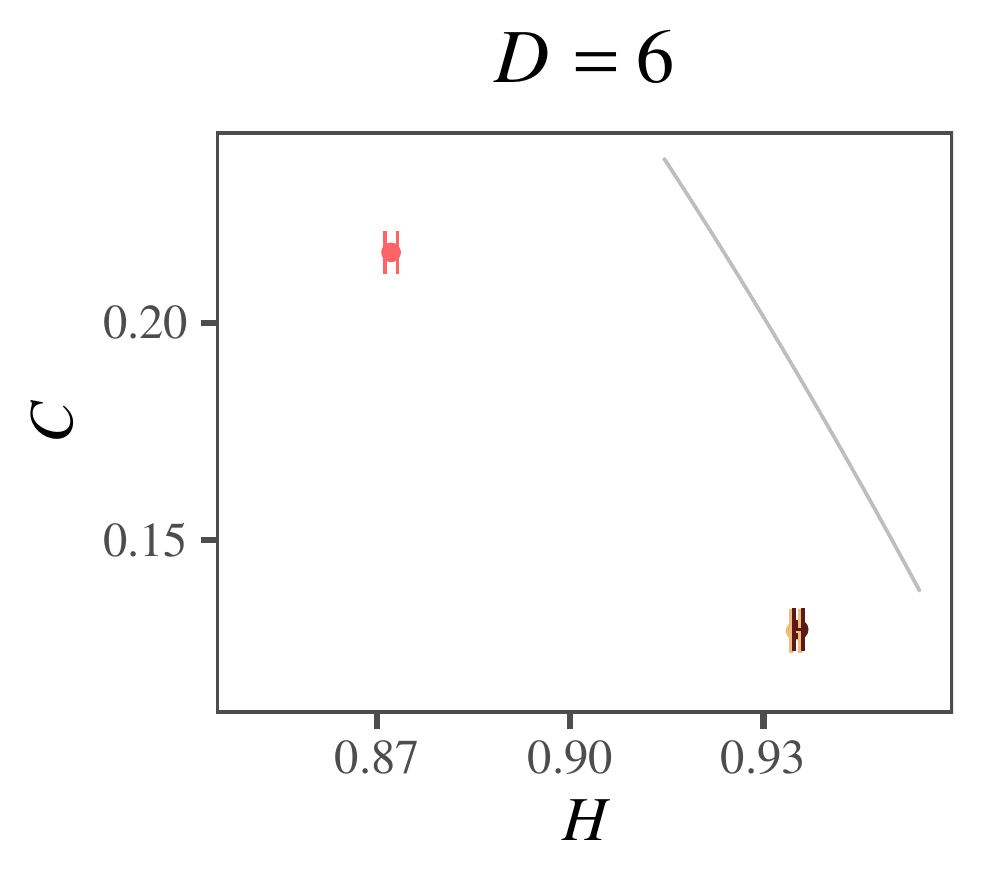}
	\caption{Points in the $H\times C$ plane from (with boundaries in light gray) the three maximum daily temperature time series, along with the proposed confidence intervals for Shannon's Entropy and different embedding dimensions $D$.}\label{Fig:HxC}
\end{figure*}

We show in Fig.~\ref{Fig:HxC} the points in the $H\times C$ plane from the three maximum daily temperature time series, along with the proposed confidence intervals at the \SI{0.1}{\percent} level for Shannon's entropy and different embedding dimensions $D\in\{3,4,5,6\}$.
The boundaries for each embedding dimension appear as light gray lines; see details about this kind of plot in the works by \citet{WhiteNoiseTestfromOrdinalPatternsintheEntropyComplexityPlane} and by \citet{GeneralizedStatisticalComplexityMeasuresGeometricalandAnalyticalProperties} and the references therein.
Notice that these are visual representations of the tests that contrast the entropies from series analyzed with the same embedding dimension, i.e., they depict the main diagonals of Table~\ref{Tab:pvalues}.

Fig.~\ref{Fig:HxC} shows that the maximum daily temperatures measured in Miami consistently have the smallest entropy and the largest complexity.
The entropy carries enough information to discriminate its underlying dynamics from those that produce the measurements in Dublin and Edinburgh.
Notice, also, that in this case, the Complexity does not add information to the problem of discriminating the underlying processes that gave rise to the temperature measurements.
The entropies from Dublin and Edinburgh are always very close, statistically indistinguishable at, approximately the \SI{95}{\percent} level of confidence in three ($D=4,5,6$) out of four cases.

Whether the observed complexity is able or not to identify different underlying dynamics in Dublin and Edinburgh is an open question since, to date, we do not have expressions for their distribution.

\section{Remarks and Conclusions}

Unlike the $\chi^2$ test, the test statistic $W$ we propose in~\eqref{Eq:TestStatisticW} does not perform a bin-by-bin comparison.
Our approach, on the one hand, represents an inevitable loss of information but, on the other hand, allows the comparison of entropies computed from different embedding dimensions.

The generality of our test promotes its application in federated learning, as it allows different sources to encode the data using different embedding.

Although we study Shannon's Entropy from ordinal patterns, the approach to obtain the asymptotic distribution is valid for the Shannon entropy computed over transition graphs and their variants.

It is noteworthy that the presented derivations do not cover the extreme cases for which $H(\bm p)=0$ or $H(\bm p)=1$.
The former does not involve any randomness.
The latter was studied by \citet{WhiteNoiseTestfromOrdinalPatternsintheEntropyComplexityPlane} using an empirical approach.

\section{Acknowledgements}\label{Sec:acknowledgements}

This work was partially funded by the CONICET, São Paulo Research Foundation (FAPESP) and National Council for Scientific and Technological Development (CNPq).

\section*{Data Availability Statement}

This work employed simulated data in Section~\ref{Sec:Numerical}.
The temperature datasets analyzed in Section~\ref{Sec:Climate} are freely available at \url{https://www.ncei.noaa.gov/cdo-web/}.

\bibliography{references}
\end{document}